\documentclass[11pt]{article}
\usepackage{amsfonts}
\usepackage{amssymb}
\usepackage{amsmath}
\usepackage{geometry}
\usepackage{graphicx}
\usepackage{color}

\def\singlespacing{\baselineskip=13pt}	
\begin{document}

\title{Effective Willis constitutive equations for periodically \\  stratified
anisotropic elastic media }
\author{A.L. Shuvalov$^{a}$, A.A. Kutsenko$^{a}$, A.N. Norris$^{a,b}$, O.
Poncelet$^{a}$ \\ \\
$^{a}\ $Universit\'{e} de Bordeaux, CNRS, UMR 5469, \\
Laboratoire de M\'{e}canique Physique, Talence 33405, France\\
$^{b}$ Mechanical and Aerospace Engineering, Rutgers University,\\
Piscataway, NJ 08854-8058, USA}
\maketitle

\begin{abstract}

A method to derive homogeneous  effective constitutive equations for periodically layered elastic media is proposed.  The crucial and novel idea underlying the procedure is that the  coefficients of the dynamic effective medium can be associated with the matrix logarithm of the propagator over a unit period.   
The  effective homogeneous equations are shown to have the structure of a Willis material, characterized by  anisotropic inertia and coupling between  momentum and strain, in addition to effective elastic constants.   Expressions are presented  
for the Willis material parameters which are formally valid at any frequency and horizontal wavenumber as long as the matrix logarithm is well defined.  
The general theory is exemplified for scalar SH motion. 
Low frequency, long wavelength expansions of the effective material parameters are also developed using a  Magnus series   and explicit estimates for the rate of convergence are derived.

\end{abstract}

\section{Introduction}

\label{sec1}

Elastic waves in periodically layered or continuous (functionally graded)
elastic media of general anisotropy have been studied extensively by
different methods. Among them is the sextic formalism of Stroh, which
incorporates the elastodynamics equations into a first-order ordinary
differential equation for the displacement-traction state vector   with 
system matrix  $\mathbf{Q}$ composed of material parameters \cite%
{Ting}. The  wave-field propagator matrix  along the stratification direction $y$, $\mathbf{M}(y,0)$,  is given by the Peano series of multiple integrals of products of $\mathbf{Q}\left(
y\right) $ \cite{Pease}. This is essentially a power series in
distance-to-wavelength ratio, which is therefore particularly well-suited
to  tackling the problem of approximating a periodically stratified medium
by an effective homogeneous medium. The Stroh formalism clarifies the
meaning of zero-order homogenization, or static averaging, of a periodic
medium by revealing  the zero-order effective material parameters 
\cite{Behrens,GN} as nothing more  than the matrix $\mathbf{Q}\left( y\right) $ 
 integrated over the period $T$,  which is  the leading term of the logarithm of
the Peano series for $\mathbf{M}(T,0)$ \cite{N}.  Static averaging implies a
non-dispersive effective medium.  Generalization to a higher-order effective
homogeneous medium, which must be dispersive, is less obvious. Its
derivation is commonly based on the long-wave dispersion of the fundamental
Bloch or Floquet branches. Their onset in arbitrary anisotropic periodically
stratified media was analyzed in \cite{N}; on this basis, the scalar-wave
equation for a transversely isotropic dispersive effective medium was
modelled in \cite{N,N1,N2} and in the subsequent literature (see e.g. \cite%
{ABDW} and its bibliography). 
A semi-analytical approach for general anisotropy   \cite{WR} (see also \cite{PBG})  fits the
long-wave Floquet dispersion to  statically averaged effective constants,
such that  the effective medium is seen as a "continuum of non-dispersive
media" that are different for different frequency and propagation direction.

{In this paper a new method is proposed for finding 
dynamic  effective constitutive equations at finite frequencies and wavelength.   This is achieved by explicit construction of effective spatially constant material coefficients that reproduce exactly the monodromy matrix $\mathbf{M}(T,0)$.  The effective constitutive theory is exact in the sense that it give  the correct displacement-traction field  at the unit-cell interfaces over arbitrary long distance of propagation.  
} Two key steps distinguish the method advocated.  First is the idea, based on the Floquet theorem, of
defining the effective medium  such that the sextic system of
elastodynamics equations in this medium has the matrix of coefficients $%
\mathbf{Q}_{\mathrm{eff}}$ equal to $i\mathbf{K,}$ where $\mathbf{K}$ is the
Floquet wave number matrix with an exact definition $i\mathbf{K}T=\ln 
\mathbf{M}(T,0).$ For the low-frequency long-wave range, a matrix logarithm $%
\ln \mathbf{M}(T,0)$ admits an expansion called the Magnus series \cite{BCOR}%
. Restricting it to the leading-order term leads to the statically averaged
effective model (see above). Taking the next-order term(s) of the Magnus
expansion for $\mathbf{K}$ reveals that, unless the variation of material
properties over a period is symmetric, the above-defined dispersive
effective medium cannot be fitted to the standard form of elastodynamic 
equations, in which the frequency dispersion and non-locality would be fully
accounted by the dependence of effective density and elasticity on $\omega $
and $k_{x}$. This motivates the second significant step in the present method
which is identifying a class of constitutive models that does fit $i\mathbf{K%
}$ with the system matrix $\mathbf{Q}_{\mathrm{eff}}$ of a homogeneous
medium. We demonstrate by construction that the model described by the
Willis constitutive relations with a dynamic stress-impulse coupling tensor  
\cite{MW,Will} provides such a class of materials. Expansions of the Willis
material coefficients based on the Magnus series for $\mathbf{K}$ are
obtained in the low-frequency long-wave range where they 
are analytic in $\omega $,~$k_{x}.$ Explicit estimates of the dependence on 
$\omega $,~$k_{x}$ are found  that enable closed-form asymptotics of
the Willis coefficients with a desired accuracy. At the same time, the definition 
$i\mathbf{K}T=\ln \mathbf{M}(T,0)$ and the fitting of the matrix $\mathbf{Q}_{%
\mathrm{eff}}\equiv i\mathbf{K}$ to the effective Willis material is
formally not restricted to the low-frequency long-wave range. 

A significant outcome  of the proposed approach is that fairly explicit expressions are obtained for the effective material parameters.  This is particularly evident for the example of SH (shear horizontal) waves discussed in detail in \S \ref{SHsec}.   In this regard the approach is  distinct from that of \cite{Will} which leads to expressions for the parameters in the (spatial) transform domain.  Note also that  the results here apply to a single realization of the layered medium, no ensemble averaging is invoked.    This point is discussed further in \S  \ref{SHsec}.

The paper proceeds as follows. The  problem is formulated
in \S \ref{sec2} where the sextic formalism for periodically stratified
media is outlined, the Floquet wave number matrix $\mathbf{K}$ introduced
and its Magnus expansion examined (see also Appendix A). The main results of
the paper are derived in \S \ref{sec3}. It starts by observing that the
matrix $i\mathbf{K}$ viewed as a sextic-system matrix $\mathbf{Q}_{\mathrm{%
eff}}$ for a homogeneous effective medium cannot be associated with a medium
from the class of anisotropic elastic materials but it does fit the Willis
model. Using the \textit{{ansatz} }that the sought effective medium is
described by the spatially homogeneous equations for a Willis material, the
corresponding system matrix $\mathbf{Q}_{\mathrm{eff}}$ is constructed and
equated with $i\mathbf{K}$. Under certain assumptions, a prescription for
unique definition of the Willis effective medium is put forward. The
remainder of \S \ref{sec3} discusses general properties of the Willis
material parameters. The example of SH wave motion in a periodic structure
is considered in \S \ref{SHsec}. It illustrates the method for defining
effective coefficients of the Willis model beyond the Magnus series
expansion (which in its turn is detailed for SH waves in Appendix B). The
obtained explicit expressions are used to solve a reflection-transmission
problem at the interface of the  effective medium. Conclusions are
presented in \S \ref{conc}.


\section{Background}

\label{sec2}

\subsection{Stroh formalism and the wave number matrix $\mathbf{K}$}

We consider a Cartesian elastic medium  with density $\rho
=\rho \left( y\right) $ and stiffness tensor $c_{ijkl}=c_{ijkl}\left(
y\right) $. Basic notations used include the superscripts $^{%
\mathrm{T}},$ $^{+}$ and $^{\ast }$ for transposition, Hermitian and complex
conjugation, respectively, and $\mathbf{T}$ for the matrix with zero
diagonal and identity off-diagonal blocks.

Taking the Fourier transforms of the equilibrium and stress-strain equations%
\begin{equation}
\sigma _{ij,i}=\rho \ddot{u}_{j},\ \ \sigma _{ij}=c_{ijkl}e_{kl},\ 
\label{0}
\end{equation}%
in all variables except $y$ leads to an ordinary differential problem for
the quasi-plane modes with the phase factor $e^{i\left( k_{x}x-\omega
t\right) },$ where $\omega $\ is the frequency and $k_{x}$ the wavenumber in
an arbitrary chosen direction $X$ orthogonal to $Y$ (rotating $X$ causes all
21 elastic constants $c_{ijkl}$  to appear). Denote the unit vectors parallel
to $X$ and $Y$ by $\mathbf{m}$ and $\mathbf{n}$ so that $x=\mathbf{m\cdot r,}
$ $y=\mathbf{n\cdot r},$ and let $\mathbf{A}\left( y\right) $ and$\ \mathbf{F%
}\left( y\right) $ be the amplitudes of displacement $\mathbf{u}$ and
traction $\mathbf{n\sigma },$ respectively. Then Eqs. (\ref{0}) combine into
the Stroh system 
\begin{equation}
\frac{\mathrm{d}}{\mathrm{d}y}\mathbf{\eta }\left( y\right) =\mathbf{Q}%
\left( y\right) \mathbf{\eta }\left( y\right)  \label{1}
\end{equation}%
for the state vector $\mathbf{\eta }$ incorporating $\mathbf{A\ }$and $%
\mathbf{F}$ \cite{Ting}. Taking it in the form $\mathbf{\eta }=\left( 
\mathbf{A,\ }i\mathbf{F}\right) ^{\mathrm{T}}$ defines the 6$\times $6
system matrix as 
\begin{equation}
\mathbf{Q}\left( y\right) =i%
\begin{pmatrix}
k_{x}\mathbf{N}_{1} & \mathbf{N}_{2} \\ 
k_{x}^{2}\mathbf{N}_{3}-\rho \omega ^{2}\mathbf{I} & k_{x}\mathbf{N}_{1}^{%
\mathrm{T}}%
\end{pmatrix}
\label{1.1}
\end{equation}%
via the 3$\times $3 blocks $\mathbf{N}_{J}$ of the Stroh matrix%
\begin{equation}
\mathbf{N}\left( y\right) =%
\begin{pmatrix}
\mathbf{N}_{1} & \mathbf{N}_{2} \\ 
\mathbf{N}_{3} & \mathbf{N}_{1}^{\mathrm{T}}%
\end{pmatrix}%
,\ 
\begin{array}{c}
\mathbf{N}_{1}=-\left( nn\right) ^{-1}\left( nm\right) =\mathbf{N}_{4}^{%
\mathrm{T}},\ \mathbf{N}_{2}=-\left( nn\right) ^{-1}, \\ 
\mathbf{N}_{3}=\left( mm\right) -\left( mn\right) \left( nn\right)
^{-1}\left( nm\right) ,%
\end{array}
\label{1.2}
\end{equation}%
which is composed of the matrices with elements $\left( nn\right)
_{jk}=n_{i}c_{ijkl}n_{l},$ $\left( nm\right) _{jk}=n_{i}c_{ijkl}m_{l}=\left(
mn\right) _{kj}$ and$\ \left( mm\right) _{jk}=m_{i}c_{ijkl}m_{l}$ (note that 
$\mathbf{N}_{2}$ is negative definite). The usual indicial symmetry of $%
c_{ijkl}$ used in (\ref{1.2}) leads to a Hamiltonian structure $\mathbf{N=TN}%
^{\mathrm{T}}\mathbf{T}$ and $\mathbf{Q=TQ}^{\mathrm{T}}\mathbf{T}$ of $%
\mathbf{N}$ and $\mathbf{Q}$. Since $\omega ,$ $k_{x}$ and $\rho ,~c_{ijkl}$
are real and hence $\mathbf{Q}\left( y\right) $ is imaginary, it follows
that $\mathbf{Q=-TQ}^{+}\mathbf{T.}\ $The latter identity on its own
suffices to ensure  energy conservation. Alternative definitions of $%
\mathbf{\eta }$ and hence of $\mathbf{Q}$ may be chosen. In general, the
eigenvalues of $\mathbf{Q}\left( y\right) $ are first-degree homogeneous
functions of $\omega ,$ $k_{x}$, which implies absence of dispersion.

Given the initial condition at some $y_{0}\left( \equiv 0\right) ,$ the
solution to (\ref{1}) is $\mathbf{\eta }\left( y\right) =\mathbf{M}\left(
y,0\right) \mathbf{\eta }\left( 0\right) ,$ where $\mathbf{M}\left(
y,0\right) $ is the 6$\times $6 matricant evaluated by the Peano series \cite%
{Pease} 
\begin{equation}
\mathbf{M}\left( y,0\right) =\mathbf{I}+\int_{0}^{y}\mathbf{Q}\left(
y_{1}\right) \mathrm{d}y_{1}+\int_{0}^{y}\mathbf{Q}\left( y_{1}\right) 
\mathrm{d}y_{1}\int_{0}^{y_{1}}\mathbf{Q}\left( y_{2}\right) \mathrm{d}%
y_{2}+\int_{0}^{y}\int_{0}^{y_{1}}\int_{0}^{y_{2}}...  \label{2}
\end{equation}%
Suppose now that $\rho ,$ $c_{ijkl}$ and hence $\mathbf{Q}$ depend on $y$
periodically with a period $T$. Denote%
\begin{equation}
y=\widetilde{y}+nT,\ \ \widetilde{y}
=y~\text{mod}\, T\in \left[ 0,T\right] ,\ 
\widetilde{\varsigma }
=\frac{ \widetilde{y}}{T}\in [ 0,1] ,
\ \ 
\left\langle \cdot \right\rangle =\frac{1}{T}\int_{0}^{T} \cdot \, \mathrm{d}%
\widetilde{y}=\int_{0}^{1}\cdot \,\mathrm{d}\widetilde{\varsigma }, 
  \label{3}
\end{equation}%
where $ \left\langle ...\right\rangle $ is the static average over a unit cell. 
{It is understood hereafter that the wave-path distance $y$ includes a large
enough number $n$ of periods, which is when the present development is of
interest. }
By (\ref{2}), the matricant $\mathbf{M}\left( T,0\right) $ over $%
\left[ 0,T\right] ,$ which is called the monodromy matrix, is%
\begin{align}
\mathbf{M}\left( T,0\right)  =\mathbf{I}+\sum\nolimits_{m=1}^{\infty }%
\mathbf{M}^{\left( m\right) }  
 =\mathbf{I}+T\left\langle \mathbf{Q}\right\rangle +T^{2}\int_{0}^{1}%
\mathbf{Q}\left( \widetilde{\varsigma }\right) \mathrm{d}\widetilde{%
\varsigma }\int_{0}^{\widetilde{\varsigma }}\mathbf{Q}\left( \widetilde{%
\varsigma }_{1}\right) \mathrm{d}\widetilde{\varsigma }_{1}+ 
\ldots
\label{2.1}
\end{align}


The wave number matrix $\mathbf{K}$ is introduced by denoting the monodromy
matrix as 
\begin{equation}
\mathbf{M}\left( T,0\right) =\exp \left( i\mathbf{K}T\right) 
\quad
\Leftrightarrow
\quad
i\mathbf{K}T=\ln \mathbf{M}\left( T,0\right) 
.  \label{4}
\end{equation}%
In the following, unless otherwise specified, $\mathbf{K}$ is understood as
defined in the first Brillouin zone, which implies the zeroth Riemann sheet
of $\ln z$ with a cut $\arg z=\pm \pi $. Using (\ref{4}), the matricant $%
\mathbf{M}\left( y,0\right) $ can be written as%
\begin{equation}
\mathbf{M}\left( y,0\right) =\mathbf{M}\left( \widetilde{y},0\right) \mathbf{%
M}\left( nT,0\right) =\mathbf{L}\left( \widetilde{y}\right) \exp \left( i%
\mathbf{K}y\right) ,  \label{5}
\end{equation}%
where $\mathbf{L}\left( \widetilde{y}\right) =\mathbf{M}\left( \widetilde{y}%
,0\right) \exp \left( -i\mathbf{K}\widetilde{y}\right) $ with $\mathbf{L}%
\left( 0\right) =\mathbf{L}\left( T\right) =\mathbf{I.}$ 
. Eq. (\ref{5})$_{2}$ represents the Floquet theorem. Denote the eigenvalues
of $\mathbf{M}\left( T,0\right) $ and $\mathbf{K}$ by $e^{iK_{\alpha }T}$
and $K_{\alpha }$ ($\alpha =1..6$), respectively. In the general case where $%
\mathbf{M}\left( T,0\right) $ and $\mathbf{K}$ have six linear independent
eigenvectors $\mathbf{w}_{\alpha },$ the Floquet theorem implies that the
wave field $\mathbf{\eta }\left( y\right) =\mathbf{M}\left( y,0\right) 
\mathbf{\eta }\left( 0\right) $ with the initial data expanded as $\mathbf{%
\eta }\left( 0\right) =\sum\nolimits_{\alpha }C_{\alpha }\mathbf{w}_{\alpha
} $ takes the form%
\begin{equation}
\mathbf{\eta }\left( y\right) =\sum\nolimits_{\alpha =1}^{6}C_{\alpha }%
\mathbf{\eta }_{\alpha }\left( y\right) \mathrm{\ where\ }\mathbf{\eta }%
_{\alpha }\left( y\right) =\mathbf{L}\left( \widetilde{y}\right) \mathbf{w}%
_{\alpha }e^{iK_{\alpha }y}.  \label{4.1}
\end{equation}

The identity $\mathbf{Q=-TQ}^{+}\mathbf{T}$ yields $\mathbf{M}^{-1}\left(
y,0\right) =\mathbf{TM}^{+}\left( y,0\right) \mathbf{T,}$ which leads in
turn to $\mathbf{L}^{-1}\left( \widetilde{y}\right) =\mathbf{TL}^{+}\left( 
\widetilde{y}\right) \mathbf{T}$ and 
\begin{equation}
\mathbf{K}=\mathbf{TK}^{+}\mathbf{T}=%
\begin{pmatrix}
\mathbf{K}_{1} & \mathbf{K}_{2} \\ 
\mathbf{K}_{3} & \mathbf{K}_{1}^{+}%
\end{pmatrix}%
\mathrm{\ with\ }\mathbf{K}_{2,3}=\mathbf{K}_{2,3}^{+}  \label{6}
\end{equation}%
for $\mathbf{K}$ defined in the first Brillouin zone. If the unit-cell
profile is symmetric, i.e. the variation of material properties within the
period $T$\ is symmetric about the middle point so that $\mathbf{Q}\left( 
\widetilde{y}\right) $ is even about $\widetilde{y}=T/2,$ then the above
identities are complemented by $\mathbf{M}\left( T,0\right) =\mathbf{TM}^{%
\mathrm{T}}\left( T,0\right) \mathbf{T}$ and hence $\mathbf{K}=\mathbf{TK}^{%
\mathrm{T}}\mathbf{T;}$ so, with reference to (\ref{6}), $\mathbf{K}$ is
real. Thus 
\begin{equation}
\mathbf{K=\mathbf{TK}}^{\mathrm{T}}\mathbf{\mathbf{T}=K}^{\ast }\ \mathrm{%
for\ a\ symmetric\ }\mathbf{Q}\left( \widetilde{y}\right) \mathrm{.}
\label{6.1}
\end{equation}

\subsection{Expansion of $\mathbf{K}$ in the Magnus series}

The logarithm of the monodromy matrix $\mathbf{M}\left( T,0\right) $ can be
expanded as a Magnus series \cite{BCOR} (see also \cite{WR4}): 
\begin{align}
i\mathbf{K}& =\frac{1}{T}\ln \mathbf{M}\left( T,0\right) =\left\langle 
\mathbf{Q}\right\rangle +\sum_{m=1}^{\infty }i\mathbf{K}^{\left( m\right) }\
\ \mathrm{with}\   \notag  \label{8} \\
i\mathbf{K}^{\left( 1\right) }& =T\frac{1}{2}\int_{0}^{1}\mathrm{d}%
\widetilde{\varsigma }\int_{0}^{\widetilde{\varsigma }}\left[ \mathbf{Q}(%
\widetilde{\varsigma }),\mathbf{Q}(\widetilde{\varsigma }_{1})\right] 
\mathrm{d}\widetilde{\varsigma }_{1}, \\
i\mathbf{K}^{\left( 2\right) }& =T^{2}\frac{1}{6}\int_{0}^{1}\mathrm{d}%
\widetilde{\varsigma }\int_{0}^{\widetilde{\varsigma }}\mathrm{d}\widetilde{%
\varsigma }_{1}\int_{0}^{\widetilde{\varsigma }_{1}}\left( \left[ \mathbf{Q,}%
\left[ \mathbf{Q,Q}\right] \right] +\left[ \left[ \mathbf{Q,Q}\right] ,%
\mathbf{Q}\right] \right) \mathrm{d}\widetilde{\varsigma }_{2},\ \mathrm{etc}%
.,  \notag
\end{align}%
where $\left[ \mathbf{Q}\left( x\right) \mathbf{,Q}\left( y\right) \right] =%
\mathbf{Q}\left( x\right) \mathbf{Q}\left( y\right) -\mathbf{Q}\left(
y\right) \mathbf{Q}\left( x\right) $ is a commutator of matrices depending
on successive integration variables. Each Magnus series term $\mathbf{K}%
^{\left( m\right) }$ is a $\left( m+1\right) $-tuple integral of
permutations of $m$ nested commutators involving products of $\left(
m+1\right) $ matrices $\mathbf{Q}\left( \widetilde{\varsigma }_{i}\right) .$
A commutator-based form may be anticipated by noting that all $\mathbf{K}%
^{\left( m\right) }$ for $m>0$ must vanish in the trivial case of a
homogeneous material with a constant $\mathbf{Q\equiv Q}_{0}$ and hence with 
$\mathbf{M}\left( T,0\right) =\exp \left( \mathbf{Q}_{0}T\right) .$ For
practical calculations  it is convenient to use the recursive formulas
provided in \cite{BCOR}. In obvious contrast with the Peano expansion, the
Magnus series converges in a limited range: the sufficient condition for its
convergence is $\left\langle \big\Vert \mathbf{Q}\big\Vert
_{2}\right\rangle <\pi /T,$ where $\big\Vert \cdot \big\Vert _{2}$ is the
matrix norm \cite{MN}. This condition implies that the eigenvalues $%
K_{\alpha }\left( \omega ^{2},k_{x}\right) $ of $\mathbf{K,}$ defined as
continuous functions such that $K_{\alpha }\left( 0,0\right) =0$, satisfy
the inequality $\Vert \text{Re}\, K_{\alpha }\Vert <\pi /T$.

The Magnus series for $\mathbf{K}$ is a low-frequency long-wave
expansion. Actually $\omega $ and $k_{x}$ are two independent parameters for 
$\mathbf{Q}$ and hence for $\mathbf{M}$ and $\mathbf{K.}$ It is however
essential that the dependence of 3$\times $3 blocks of $\mathbf{Q}$ on $%
\omega $ and $k_{x}$ is homogeneous, see (\ref{1.1}). Therefore the blocks
of each $m$th term of the Peano and Magnus series are homogeneous
polynomials of $\omega $, $k_{x}$\ of degree  one greater than the same
block of the $\left( m-1\right) $th term. This is what enables introducing a
single long-wave parameter $\varepsilon \equiv kT$ with a suitably defined
wavenumber $k$, see \eqref{7}\textit{.} Thus the Magnus series (\ref{8}) is
basically an expansion in powers of $\varepsilon $.
%
Taking small enough $\varepsilon $ enables its approximation by a finite
number of terms. At the same time, it should be borne in mind that the
Magnus series as an expansion of logarithm may converge relatively slowly.
Explicit estimates expressed in terms of $\omega $, $k_{x}$ and $%
\left\langle \mathbf{N}\right\rangle $ which ensure a desired accuracy of
truncating the Magnus series up to a given order are formulated in Appendix
A.

The structure of polynomial dependence of the Magnus series terms $\mathbf{K}%
^{\left( m\right) }$ on $k_{x}$ and $\omega ^{2}$ is 
\begin{align}
\mathbf{K}^{\left( 1\right) }& =i%
\begin{pmatrix}
k_{x}^{2}\mathbf{a}_{1}^{\left( 1\right) }+\omega ^{2}\mathbf{a}_{2}^{\left(
1\right) } & k_{x}\mathbf{a}_{3}^{\left( 1\right) } \\ 
k_{x}(k_{x}^{2}\mathbf{a}_{4}^{\left( 1\right) }+\omega ^{2}\mathbf{a}%
_{5}^{\left( 1\right) }) & -k_{x}^{2}\mathbf{a}_{1}^{\left( 1\right) \mathrm{%
T}}-\omega ^{2}\mathbf{a}_{2}^{\left( 1\right) }%
\end{pmatrix}%
,  \notag  \label{9.2} \\
\mathbf{K}^{\left( 2\right) }& =%
\begin{pmatrix}
k_{x}(k_{x}^{2}\mathbf{a}_{1}^{\left( 2\right) }+\omega ^{2}\mathbf{a}%
_{2}^{\left( 2\right) }) & k_{x}^{2}\mathbf{a}_{3}^{\left( 2\right) }+\omega
^{2}\mathbf{a}_{4}^{\left( 2\right) } \\ 
k_{x}^{2}(k_{x}^{2}\mathbf{a}_{5}^{\left( 2\right) }+\omega ^{2}\mathbf{a}%
_{6}^{(2)})+\omega ^{4}\mathbf{a}_{7}^{(2)} & k_{x}(k_{x}^{2}\mathbf{a}%
_{1}^{\left( 2\right) \mathrm{T}}+\omega ^{2}\mathbf{a}_{2}^{\left( 2\right) 
\mathrm{T}})%
\end{pmatrix}%
, \\
\mathbf{K}^{\left( 3\right) }& =i%
\begin{pmatrix}
k_{x}^{2}(k_{x}^{2}\mathbf{a}_{1}^{\left( 3\right) }+\omega ^{2}\mathbf{a}%
_{2}^{\left( 3\right) })+\omega ^{4}\mathbf{a}_{3}^{\left( 3\right) } & 
k_{x}(k_{x}^{2}\mathbf{a}_{4}^{\left( 3\right) }+\omega ^{2}\mathbf{a}%
_{5}^{\left( 3\right) }) \\ 
k_{x}(k_{x}^{4}\mathbf{a}_{6}^{(3)}+\omega ^{2}k_{x}^{2}\mathbf{a}%
_{7}^{\left( 3\right) }+\omega ^{4}\mathbf{a}_{8}^{\left( 3\right) }) & 
-k_{x}^{2}(k_{x}^{2}\mathbf{a}_{1}^{\left( 3\right) \mathrm{T}}+\omega ^{2}%
\mathbf{a}_{2}^{\left( 3\right) \mathrm{T}})-\omega ^{4}\mathbf{a}%
_{3}^{\left( 3\right) \mathrm{T}}%
\end{pmatrix}%
\ \mathrm{etc.},  \notag
\end{align}%
where the real matrices $\mathbf{a}^{\left( m\right) }$ in $\mathbf{K}%
^{\left( m\right) }$ are $\left( m+1\right) $-tuple integrals of appropriate
commutators; for instance, $\mathbf{a}_{i}^{\left( 1\right) }$ in $\mathbf{K}%
^{\left( 1\right) }$ are 
\begin{align}
 \big\{ 
\mathbf{a}_{1}^{( 1)},\ 
\mathbf{a}_{2}^{( 1)},\ 
\mathbf{a}_{3}^{( 1)},\ 
\mathbf{a}_{4}^{( 1)},\ 
\mathbf{a}_{5}^{( 1)} \big\} 
&
=\frac{1}{2} T
\int_{0}^{1}\mathrm{d}\widetilde{\varsigma } 
\int_{0}^{\widetilde{\varsigma }}  \mathrm{d}\widetilde{\varsigma }_{1} 
 \bigg\{  
\left[ \mathbf{N}_{1},\mathbf{N}_{1}\right] +\left[ \mathbf{N}_{2},\mathbf{N}_{3}\right]
,\ 
\left[ \rho \mathbf{I},\mathbf{N}_{2}\right]  
,\ 
  \notag  \label{9.3} 
\\
 &
 \left[ \mathbf{N}_{1},\mathbf{N}_{2}\right] +\left[ \mathbf{N}_{2},\mathbf{N}_{1}^{%
\mathrm{T}}\right] 
, \ \left[ \mathbf{N}_{3},\mathbf{N}_{1}\right] +\left[ \mathbf{N%
}_{1}^{\mathrm{T}},\mathbf{N}_{3}\right] 
,\ \left[ \mathbf{N%
}_{1}-\mathbf{N}_{1}^{\mathrm{T}},\rho \mathbf{I}\right] \bigg\} .
\end{align}%
The series terms $\mathbf{K}^{\left( m\right) }$ of odd and even order $m$
are imaginary and real, respectively, and each term $\mathbf{K}^{\left(
m\right) }$ on its own satisfies (\ref{6}); therefore%
\begin{equation}
\mathbf{K}^{\left( m\right) }=-\mathbf{K}^{\left( m\right) \ast }=-\mathbf{TK%
}^{\left( m\right) \mathrm{T}}\mathbf{T\ }\mathrm{for\ odd}\mathbf{\ }m,\ 
\mathbf{\ K}^{\left( m\right) }=\mathbf{K}^{\left( m\right) \ast }=\mathbf{TK%
}^{\left( m\right) \mathrm{T}}\mathbf{T}\mathrm{\ for}\mathbf{\ }\mathrm{even%
}\mathbf{\ }m,  \label{9}
\end{equation}%
as taken into account in (\ref{9.2}). According to (\ref{6.1}),%
\begin{equation}
\mathbf{K}^{\left( m\right) }=\mathbf{0\ }\mathrm{for\ odd}\ m,\ \mathrm{if}%
\ \mathbf{Q}\left( \widetilde{y}\right) \mathrm{\ is\ symmetric.}
\label{9.1}
\end{equation}%
Note the pure dynamic imaginary terms proportional to $\pm i\omega ^{m+1},$
which appear in the diagonal blocks of the series terms $\mathbf{K}^{\left(
m\right) }$ of odd order $m$ unless these are zero for a symmetric $\mathbf{Q%
}\left( \widetilde{y}\right) $ by (\ref{9.1}).

\subsection{Dynamic homogenization}

According to the Floquet theorem (\ref{5}), the wave field variation over a
large distance $y$ is characterized mainly by the function $\exp \left( i%
\mathbf{K}y\right) $ (which is an exact wave field at $y=nT$). Formally $%
\exp \left( i\mathbf{K}y\right) $ with $i\mathbf{K}T=\ln \mathbf{M}\left(
T,0\right) $ is a solution to Eq. (\ref{1}) with the actual matrix of
coefficients $\mathbf{Q}\left( y\right) $ replaced by the constant matrix $i%
\mathbf{K.}$ 
This motivates the concept of an effective homogeneous medium, whose
material model admits the wave equation in the form (\ref{1}) with a
constant system matrix $\mathbf{Q}_{\mathrm{eff}}\equiv i\mathbf{K}$.
Confining the Magnus series (\ref{8}) to the zero-order term   defines the
statically averaged $\mathbf{Q}_{\mathrm{eff}}^{\left( 0\right)
}=\langle \mathbf{Q}\rangle $ which fits (\ref{1.1}) with $%
\mathbf{N}_{\mathrm{eff}}=\left\langle \mathbf{N}\right\rangle $ and hence
yields the non-dispersive effective density and stiffness in the well-known
form $\rho ^{\left( 0\right) }=\left\langle \rho \right\rangle $ and 
\begin{equation}
\left( nn\right) ^{\left( 0\right) }=-\left\langle \mathbf{N}%
_{2}\right\rangle ^{-1},\ \left( nm\right) ^{\left( 0\right) }=\left\langle 
\mathbf{N}_{2}\right\rangle ^{-1}\left\langle \mathbf{N}_{1}\right\rangle ,\
\left( mm\right) ^{\left( 0\right) }=\left\langle \mathbf{N}%
_{3}\right\rangle -\left\langle \mathbf{N}_{1}^{\mathrm{T}}\right\rangle
\left\langle \mathbf{N}_{2}\right\rangle ^{-1}\left\langle \mathbf{N}%
_{1}\right\rangle ,  \label{10}
\end{equation}%
see \cite{N}. It is evident that the statically averaged $\mathbf{Q}_{\mathrm{eff%
}}^{\left( 0\right) }=\left\langle \mathbf{Q}\right\rangle $ completely
ignores dynamic effects and is  inadequate to describe waves at
finite frequency over long propagation distance. Dynamic properties
are realized by taking $\mathbf{Q}_{\mathrm{eff}}=i\mathbf{K}$ beyond the
zero-order term $\left\langle \mathbf{Q}\right\rangle ,$ see next Section. 
Note that, in contrast to $\langle \mathbf{Q}\rangle $, a dispersive 
$\mathbf{Q}_{\mathrm{eff}}=i \mathbf{K}$ 
 generally depends on where the reference point $y=0$ of the period interval $[0,T]$ is chosen. 

\section{A dispersive effective medium with $\mathbf{Q}_{\mathrm{eff}}=i%
\mathbf{K}$}

\label{sec3}

\subsection{The constitutive equations}

Our purpose is to take into account the full nature of the wave number
matrix in $\mathbf{Q}_{\mathrm{eff}}=i\mathbf{K}$. With this in mind,
compare the structure of the dispersive effective matrix as given by the
Magnus expansion $\mathbf{Q}_{\mathrm{eff}}=\mathbf{\mathbf{\left\langle 
\mathbf{Q}\right\rangle }}+i\sum_{m=1}\mathbf{K}^{\left( m\right) }$ with
that of $\mathbf{Q}\left( y\right) $ given by (\ref{1.1}). They differ in
two ways. First, $\mathbf{Q}_{\mathrm{eff}}$ ($\neq \mathbf{\mathbf{%
\left\langle \mathbf{Q}\right\rangle }}$) is no longer imaginary, and hence
the identity $\mathbf{Q}_{\mathrm{eff}}=\mathbf{-TQ}_{\mathrm{eff}}^{+}%
\mathbf{T}$ which leads to (\ref{6}) (and ensures energy conservation) is no
longer compatible with a Hamiltonian structure for $\mathbf{Q}_{\mathrm{eff}%
}.$ This is a well-known feature of dispersive models, see e.g. \cite{PB}.
The second, more significant, dissimilarity is that, by contrast to (\ref%
{1.1}), $\mathbf{Q}_{\mathrm{eff}}$ has pure dynamic terms on the diagonal,
already at the first-order $i\mathbf{K}^{\left( 1\right) },$ see (\ref{9.2}%
). This means that assuming dispersive density and elastic constants does
not  allow the constitutive relations of the dispersive effective medium
to be written in the standard form of equations (\ref{0}). Recalling that
the upper rows of the sextic system (\ref{1}) imply the traction-strain law,
it is seen that the latter must be complemented by a purely dynamic term
which implies different constitutive relations than those of the
inhomogeneous medium itself.

On this basis, following \cite{MW,Will}, the equations of equilibrium and the
constitutive relations \eqref{0} are replaced by the more general form
proposed by Willis%
\begin{equation}
\sigma _{ij,i}=\dot{p}_j,\ \ \sigma _{ij}=c_{ijkl}^{( \mathrm{eff} )) }e_{kl}+S_{ijr}\dot{u}_{r},\ \ p_{q}=S_{klq}e_{kl}+\rho
_{qr}^{( \mathrm{eff} ) }\dot{u}_{r}.  \label{11}
\end{equation}%
The vector $\mathbf{p}$ generalizes the normal notion of momentum density,
and the elements of the Willis coupling tensor {satisfy $S_{ijk}= S_{jik}$
by assumption, ensuring the symmetry of the stress tensor}. A principal
objective is to show that setting $\mathbf{Q}_{\mathrm{eff}}=i\mathbf{K}$
leads inevitably to dispersive effective matrix density $\pmb{\rho }%
^{( \mathrm{eff} ) }$ and stiffness $c_{ijkl}^{( \mathrm{eff} )) }$ and, on top of that, to the Willis form of the effective
constitutive relations with stress-impulse coupling.


\subsection{The effective Willis medium}

Denote by $\mathbf{S}_{\mathbf{n}}$ and $\mathbf{S}_{\mathbf{m}}\ $the
matrices with components 
\begin{equation}
\left( \mathbf{S}_{\mathbf{n}}\right) _{jk}=n_{i}S_{ijk},\ \left( \mathbf{S}%
_{\mathbf{m}}\right) _{jk}=m_{i}S_{ijk}.  \label{11.1}
\end{equation}%
The same derivation that led from (\ref{0}) to the sextic system (\ref{1})
with the coefficients (\ref{1.1}) now leads from (\ref{11}) to (\ref{1})
with the system matrix%
\begin{equation}
\mathbf{Q}_{\mathrm{eff}} =i%
\begin{pmatrix}
k_{x}\mathbf{N}_{1}^{( \mathrm{eff} ) }-\omega \mathbf{N}%
_{2}^{( \mathrm{eff} ) }\mathbf{S}_{\mathbf{n}} & \mathbf{N}%
_{2}^{( \mathrm{eff} ) } \\ 
k_{x}^{2}\mathbf{N}_{3}^{( \mathrm{eff} ) }-\omega ^{2}(\mathbf{%
\rho }^{( \mathrm{eff} ) }-\mathbf{S}_{\mathbf{n}}^{+}\mathbf{N}%
_{2}^{( \mathrm{eff} ) }\mathbf{S}_{\mathbf{n}})-\omega k_{x}%
\mathbf{L} & ~k_{x}\mathbf{N}_{1}^{( \mathrm{eff} ) +}-\omega 
\mathbf{S}_{\mathbf{n}}^{+}\mathbf{N}_{2}^{(\mathrm{eff})}%
\end{pmatrix}%
,\   \label{12} 
\end{equation}
with $\mathbf{L}=
\mathbf{S}_{\mathbf{n}}^{+}\mathbf{N}%
_{1}^{( \mathrm{eff}) }+\mathbf{N}_{1}^{( \mathrm{eff}%
) +}\mathbf{S}_{\mathbf{n}}+\mathbf{S}_{\mathbf{m}}+\mathbf{S}_{%
\mathbf{m}}^{+}$ $=\mathbf{L}^{+}$. 
The identity $\mathbf{Q}_{\mathrm{eff}}=\mathbf{-TQ}_{\mathrm{eff}}^{+}\mathbf{T}$  is assumed in order to ensure that the effective medium is, like the inhomogeneous periodic medium,   non-dissipative (energy conserving).  This 
implies hermiticity constraints on the complex-valued material parameters: $%
\mathbf{c}^{( \mathrm{eff}) }=\mathbf{c}^{( \mathrm{eff}%
) +}$, $\pmb{\rho }^{( \mathrm{eff}) }=\pmb{\rho }%
^{( \mathrm{eff} ) +}$, $\mathbf{S}_{\mathbf{n}}=-\mathbf{S}_{%
\mathbf{n}}^{\ast }$, $\mathbf{S}_{\mathbf{m}}=-\mathbf{S}_{\mathbf{m}%
}^{\ast }$, where $\mathbf{c}^{( \mathrm{eff} ) }$ is the 6$\times 
$6 stiffness matrix in Voigt's notation. The coupling tensor $S_{ijk}$ is
therefore purely imaginary. Note that the blocks $\mathbf{N}_{J}^{\left( 
\mathrm{eff}\right) }$ of the effective Stroh matrix $\mathbf{N}_{\mathrm{eff%
}}=\mathbf{TN}_{\mathrm{eff}}^{+}\mathbf{T}$ $(\neq \mathbf{TN}_{\mathrm{eff}%
}^{\mathrm{T}}\mathbf{T})$ consist of sub-matrices $\left( nn\right)
^{( \mathrm{eff} ) },~\left( nm\right) ^{( \mathrm{eff} )) },$ $\left( mm\right) ^{( \mathrm{eff} ) }$ built from $%
\mathbf{c}^{( \mathrm{eff} ) }$ according to the definition (\ref%
{1.2}) but with $\left( mn\right) ^{( \mathrm{eff} ) }=\left(
nm\right) ^{( \mathrm{eff} ) +}$ so that $\mathbf{N}_{4}^{\left( 
\mathrm{eff}\right) }=\mathbf{N}_{1}^{( \mathrm{eff} ) +}$.

Equating the matrix $\mathbf{Q}_{\mathrm{eff}}$ introduced in (\ref{12}) to
the matrix $i\mathbf{K}\left( \omega ^{2},k_{x}\right) $ with the block
structure \eqref{6} yields the blockwise equalities 
\begin{align} & \quad
k_{x}\mathbf{N}_{1}^{( \mathrm{eff} ) }-\omega \mathbf{N}%
_{2}^{( \mathrm{eff} ) }\mathbf{S}_{\mathbf{n}} =\mathbf{K}_{1}, 
\quad \mathbf{N}_{2}^{( \mathrm{eff} ) } =\mathbf{K}_{2},  
\notag   \\
&
k_{x}^{2}\mathbf{N}_{3}^{( \mathrm{eff} ) }-\omega ^{2}(\mathbf{%
\rho }^{( \mathrm{eff} ) }-\mathbf{S}_{\mathbf{n}}^{+}\mathbf{N}%
_{2}^{( \mathrm{eff} ) }\mathbf{S}_{\mathbf{n}})-\omega k_{x}%
\mathbf{L} =\mathbf{K}_{3}.    \label{10.1}
\end{align}%
Identification of the effective parameters based on these identities is
ambiguous given that they may depend upon both $\omega $ and $k_{x}$. A
unique and arguably the simplest solution is obtained by first assuming a
purely dynamic $S_{ijk}=S_{ijk}\left( \omega \right) $. Then (\ref{10.1}$%
_{1} $) and (\ref{10.1}$_{2}$) yield%
\begin{align}
\omega \mathbf{S}_{\mathbf{n}} =-\mathbf{K}_{2}^{-1}\left( 0\right) \mathbf{%
K}_{1}\left( 0\right) ,  
\quad 
\left( nn\right) ^{( \mathrm{eff} ) } =-\mathbf{K}_{2}^{-1}, 
\quad
k_{x}\left( nm\right) ^{( \mathrm{eff} ) } =\mathbf{K}_{2}^{-1}%
\mathbf{K}_{1}-\mathbf{K}_{2}^{-1}\left( 0\right) \mathbf{K}_{1}\left(
0\right) ,  \   \label{10.2}  
\end{align}%
where the blocks $\mathbf{K}_{J}\left( \omega ^{2},k_{x}\right) $ of $%
\mathbf{K}$ are taken at $k_{x}=0$ as indicated by the notation $\mathbf{K}%
_{J}\left( 0\right) \equiv \mathbf{K}_{J}\left( \omega ^{2},0\right) $ used
here and subsequently. Note that the validity of \eqref{10.2} requires the
block $\mathbf{K}_{2}$ to be invertible which is assumed in the following.
For $\mathbf{K}$ defined by the Magnus series (\ref{8}), the conditions that
ensure existence of $\mathbf{K}_{2}^{-1}$ in a certain low-frequency
long-wave range and the estimates that enable truncation of its expansion
are established in Appendix A. 
In view of \eqref{10.2}, the remaining identity (\ref{10.1}$_{3}$) becomes 
\begin{equation}
k_{x}^{2}\left( mm\right) ^{( \mathrm{eff} ) }-\omega ^{2}\mathbf{%
\rho }^{( \mathrm{eff} ) }-\omega k_{x}\left( \mathbf{S}_{\mathbf{m%
}}+\mathbf{S}_{\mathbf{m}}^{+}\right) =\mathbf{K}_{3}-\mathbf{K}_{1}^{+}%
\mathbf{K}_{2}^{-1}\mathbf{K}_{1}.  \label{10.2d}
\end{equation}%
Pursuing the logical extension of the assumed dependence of the inertial
coupling tensor $\mathbf{S}$ on frequency alone, we assume that the inertia
tensor $\pmb{\rho }^{( \mathrm{eff} ) }$ is also purely
dynamic. This leads to a unique solution of \eqref{10.2d} since $\mathbf{%
\rho }^{( \mathrm{eff} ) }=\pmb{\rho }^{( \mathrm{eff} )) }(\omega )$ is found by setting $k_{x}=0$, 
\begin{equation}
-\omega (\mathbf{S}_{\mathbf{m}}+\mathbf{S}_{\mathbf{m}}^{+})=\lim_{k_{x}%
\rightarrow 0}\,k_{x}^{-1}\big\{\mathbf{K}_{3}-\mathbf{K}_{1}^{+}\mathbf{K}%
_{2}^{-1}\mathbf{K}_{1}-\big(\mathbf{K}_{3}(0)-\mathbf{K}_{1}^{+}(0)\mathbf{K%
}_{2}^{-1}(0)\mathbf{K}_{1}(0)\big)\big\}.
\end{equation}%
The limit may be achieved in terms of derivatives of matrices $\mathbf{K}%
_{J} $ at $k_{x}=0$, whose existence is guaranteed for instance within the
range of convergence of the Magnus expansion. Accordingly, the solutions of %
\eqref{10.2d} are  
\begin{align}
\omega ^{2}\pmb{\rho }^{( \mathrm{eff} ) }& =\mathbf{K}%
_{1}^{+}(0)\mathbf{K}_{2}^{-1}(0)\mathbf{K}_{1}(0)-\mathbf{K}_{3}(0),  \notag
\label{-1} \\
\omega (\mathbf{S}_{\mathbf{m}}+\mathbf{S}_{\mathbf{m}}^{+})& =-\mathbf{K}%
_{3}^{\prime }\left( 0\right) -\omega \mathbf{S}_{\mathbf{n}}^{+}\mathbf{K}%
_{1}^{\prime }\left( 0\right) -\omega \mathbf{K}_{1}^{+\prime }\left(
0\right) \mathbf{S}_{\mathbf{n}}-\omega ^{2}\mathbf{S}_{\mathbf{n}}^{+}%
\mathbf{K}_{2}^{\prime }\left( 0\right) \mathbf{S}_{\mathbf{n}}, \\
k_{x}^{2}\left( mm\right) ^{( \mathrm{eff} ) }& =\mathbf{K}_{3}-%
\mathbf{K}_{1}^{+}\mathbf{K}_{2}^{-1}\mathbf{K}_{1}+\omega ^{2}\pmb{\rho }%
^{( \mathrm{eff} ) }+\omega k_{x}\left( \mathbf{S}_{\mathbf{m}}+%
\mathbf{S}_{\mathbf{m}}^{+}\right) .  \notag
\end{align}%
 where $\mathbf{K}_{J}^{\prime }\left( 0\right) = 
\partial \mathbf{K}
_{J}\left( \omega ^{2},k_{x}\right) /\partial k_{x}\big\Vert _{k_{x}=0}.$

In summary, Eqs. (\ref{10.2}) and \eqref{-1} provide unique material
properties for the Willis model with 
\begin{equation}
\mathbf{c}^{( \mathrm{eff} ) }(\omega ,k_{x})=\mathbf{c}^{\left( 
\mathrm{eff}\right) +},\ \pmb{\rho }^{( \mathrm{eff} ) }(\omega
)=\pmb{\rho }^{( \mathrm{eff} ) +},\ \mathbf{S}(\omega )=-%
\mathbf{S}^{\ast },  \label{12.1}
\end{equation}%
and $S_{ijk}=S_{jik}$. The lack of dependence of the inertial parameters on $%
k_{x}$ means that non-local effects are confined to the elastic moduli $%
\mathbf{c}^{( \mathrm{eff} ) }$. The result reduces to
non-dispersive statically averaged moduli of (\ref{10}) with $\pmb{\rho }%
^{( \mathrm{eff} ) }=\langle \rho \rangle \mathbf{I}$ and $\mathbf{%
S=0}$ when $i\mathbf{K}$ is restricted to the zero-order term $\left\langle 
\mathbf{Q}\right\rangle $ of (\ref{8}).

Regarding computation of the Willis parameters from Eqs. \eqref{10.2} and %
\eqref{-1}, it is assumed that the wavenumber matrix $\mathbf{K}\left(
\omega ^{2},k_{x}\right) $ defined as $i\mathbf{K}T=\ln \mathbf{M}\left(
T,0\right) $ is known either in the form of long-wave low-frequency series,
or from the direct definition of matrix logarithm in some neighbourhood of a
given point $\omega $, $k_{x}$ (see the example in \S \ref{SHsec}). In
particular, one may first evaluate the matricant $\mathbf{M}\left(
T,0\right) $ numerically and $\mathbf{K}$ then follows from the matrix
logarithm. The matrix $\mathbf{K}^{\prime }(0)$, required for the solution
of Eq. \eqref{-1}$_{2}$, involves evaluating the derivative of $\ln \mathbf{M%
}\left( T,0\right) $ in $k_{x}.$ It may be expressed using
either series or integral representation of $\ln \mathbf{M}$ as%
\begin{equation}
\ln \mathbf{M=-}\sum\nolimits_{n=1}^{\infty }\frac{1}{n}\left( \mathbf{I}-%
\mathbf{M}\right) ^{n}=( \mathbf{M-I}) 
\int_{0}^{1 } [ x( \mathbf{M} -\mathbf{I}) +\mathbf{I}]^{-1}\, 
\mathrm{d}x,  \label{++1}
\end{equation}%
which leads to 
\begin{equation}
i\mathbf{K}^{\prime }(0)T
=\sum\limits_{n=1}^{\infty } \frac 1 n (-1)^{n-1}  
\sum\limits_{j=0}^{n-1}  \mathbf{A}^{j}%
\mathbf{M}^{\prime }(0)  \mathbf{A} ^{n-1-j} 
 =
\int_{0}^{1 } [  x  \mathbf{A}+\mathbf{I}]^{-1}
\mathbf{M}^{\prime}(0)
[  x  \mathbf{A} +\mathbf{I}]^{-1}
\mathrm{d}x,%
\label{++2}
\end{equation}
with  $\mathbf{A}= \mathbf{M}\left( T,0\right)-\mathbf{I} $ 
at $%
k_{x}=0,$ and the derivative of the matricant itself is \cite{Pease}%
\begin{equation}
\mathbf{M}^{\prime }(0)\equiv   \frac{\partial \mathbf{M}(T,0)}{%
\partial k_{x}}\bigg\vert _{k_{x}=0}=i\int_{0}^{T}\mathbf{M}(T,\widetilde{%
\varsigma })%
\begin{pmatrix}
\mathbf{N}_{1} & 0 \\ 
0 & \mathbf{N}_{1}^{\mathrm{T}}%
\end{pmatrix}%
\mathbf{M}(\widetilde{\varsigma },0)\mathrm{d}\widetilde{\varsigma }.\ \ 
\label{++}
\end{equation}%
The sufficient conditions for the range of validity of the above series and
integral definitions of $\mathbf{K}^{\prime }(0)$ are specified in Appendix
A.

\subsection{Discussion}

\subsubsection{The Willis equation and its inertial quantities}

\label{dis1}

Consider the above results \eqref{10.2} and \eqref{-1} in more detail.
Anisotropic density and the coupling coefficients that relate particle
momentum and stress are unknown in "standard" models of solids. Here they
appear as inevitable ingredients of a model that replaces periodic spatial
inhomogeneity with a spatially homogeneous but dispersive and non-local
theory. The departure from normal elasticity is evident from the equations
of motion for the displacement that follows from \eqref{11}, 
\begin{equation}
c_{ijkl}^{( \mathrm{eff} ) }u_{l,ik}+(S_{ijl}-S_{ilj})\dot{u}%
_{l,i}-\rho _{jl}^{( \mathrm{eff} ) }\ddot{u}_{l}=0.  \label{-2}
\end{equation}%
This in turn leads to an energy conservation equation of the form $\dot{U}+%
\mathrm{div}\,\mathbf{f}=0$ where the real-valued energy density and flux
vector are 
\begin{equation}
U=\frac{1}{2}c_{ijkl}^{( \mathrm{eff} ) }u_{l,k}u_{j,i}^{\ast }+%
\frac{1}{2}\rho _{jl}^{( \mathrm{eff} ) }\dot{u}_{l}\dot{u}%
_{j}^{\ast },\quad f_{i}=-\frac{1}{2}(S_{ijl}-S_{ilj})\dot{u}_{l}\dot{u}%
_{j}^{\ast }-\mathrm{Re}\big(c_{ijkl}^{( \mathrm{eff} ) }u_{l,k}%
\dot{u}_{j}^{\ast }\big).  \label{-2-}
\end{equation}

In order to gain some insight into these new dynamic terms, consider a
layered transversely isotropic medium, with the principal axis along $%
\mathbf{n}$ identified as the 2-direction. The effective density is of the
form $\pmb{\rho }^{( \mathrm{eff} ) }=$ diag$(\rho
_{11}^{( \mathrm{eff} ) },\rho _{22}^{( \mathrm{eff} )
},\rho _{11}^{( \mathrm{eff} ) })$ and the only non-zero elements
of the coupling tensor (up to symmetries $S_{ijk}=S_{jik}$) are $%
S_{112}=S_{332}$, $S_{211}=S_{233}$ and $S_{222}$. Only one combination of
the three independent coupling elements has impact on the equations of
motion, 
\begin{equation}
\begin{array}{l}
c_{ijkl}^{( \mathrm{eff} ) }u_{l,ik}+(S_{112}-S_{211})\dot{u}%
_{2,j}-\rho _{11}^{( \mathrm{eff} ) }\ddot{u}_{j}=0,\ \ j=1,3, \\ 
\\ 
c_{i2kl}^{( \mathrm{eff} ) }u_{l,ik}
+(S_{211}-S_{112})( \dot{u}_{1,1}+\dot{u}_{3,3}) 
-\rho _{22}^{\left( \mathrm{eff%
}\right) }\ddot{u}_{2}=0,%
\end{array}
\label{-24}
\end{equation}%
and it has no influence on pure SH wave motion (polarized in the plane
orthogonal to $\mathbf{n}$). Long-wave expansions of $\rho _{jj}^{\left( 
\mathrm{eff}\right) }$ and $S_{ijk}$ are presented in \eqref{10.3}. Further
detailed discussion for SH waves is provided in \S \ref{SHsec}.

More generally, the absence of generating functions for $\mathbf{S}_{\mathbf{%
m}}-\mathbf{S}_{\mathbf{m}}^{+}$ means that some elements of the Willis
coupling tensor $S_{ijk}$ should be set to zero in order to complete its
definition. The relevant elements are necessary to determine stress and
momentum but do not enter into the equation of motion \eqref{-2} and the
sextic system (\ref{1}) with (\ref{12}) because the purely imaginary
property of the coupling tensor means that $m_{i}(S_{ijl}-S_{ilj})$ are the
elements of $\mathbf{S}_{\mathbf{m}}+\mathbf{S}_{\mathbf{m}}^{+}$. Consider
the two dimensional situation with indices taking only two values so that,
on account of the symmetry $S_{ijk}=S_{jik},$ there are at most six
independent elements. Four of these may be found from \eqref{10.2}$_{1}$,
and one more follows from \eqref{-1}$_{2}$ using the symmetry property. The
single element $\mathbf{m}\cdot \mathbf{S}_{\mathbf{m}}\mathbf{m}$ is
undefined and may be set equal to zero. In the three dimensional situation
all but four combinations of the 18 independent elements of $S_{ijk}$ are
obtainable. Taking an orthonormal triad $\{\mathbf{m}_{1},\mathbf{n},\mathbf{%
m}_{2}\}$, the following elements of the coupling tensor are not defined by
the effective medium equations and are therefore set to zero: $\mathbf{m}%
_{\alpha }\cdot \mathbf{S}_{\mathbf{m}_{\alpha }}\mathbf{m}_{\beta }+\mathbf{%
m}_{\beta }\cdot \mathbf{S}_{\mathbf{m}_{\alpha }}\mathbf{m}_{\alpha }$, $%
\alpha ,\beta \in \{1,2\}$. To be explicit, let $\mathbf{n}$ lie in the
2-direction, then the $\mathbf{S}_{\mathbf{n}}$ equation \eqref{10.2}$_{1}$
defines the nine elements $S_{2jk}$; these combined with the $\mathbf{S}_{%
\mathbf{m}}+\mathbf{S}_{\mathbf{m}}^{+}$ equations \eqref{-1}$_{2}$ yield $%
S_{112}$, $S_{132}$, $S_{32}$, and the $\mathbf{S}_{\mathbf{m}}+\mathbf{S}_{%
\mathbf{m}}^{+}$ equations with the above prescriptions give $%
S_{113}=-S_{131}$, $S_{313}=-S_{331}$, $S_{111}=0$, $S_{333}=0$.

\subsubsection{Expansion of the Willis parameters}

\label{expan}

Explicit insight into the structure of the Willis parameters can be gained
from their expansion obtained via the Magnus series for the wave number
matrix $\mathbf{K}$. In view of (\ref{9.2}) and (\ref{9}), $\mathbf{K}%
_{1}\left( 0\right) $ is imaginary and expands as $\mathbf{K}_{1}\left(
0\right) =\sum_{m}\mathbf{K}_{1}^{\left( m\right) }\left( 0\right) $ with
odd $m$ and $\mathbf{K}_{1}^{\left( m\right) }\left( 0\right) \sim i\omega
^{m+1},$ while $\mathbf{K}_{2}\left( 0\right) $ is real and expands as $%
\mathbf{K}_{2}\left( 0\right) =\left\langle \mathbf{N}_{2}\right\rangle
+\sum_{m}\mathbf{K}_{2}^{\left( m\right) }\left( 0\right) $ with even $m$
and $\mathbf{K}_{2}^{\left( m\right) }\left( 0\right) \sim \omega ^{m}$.
This confirms that $\mathbf{S}_{\mathbf{n}}$ is imaginary and vanishes at $%
\omega =0.$ It is easy to check that the right-hand sides of (\ref{10.2}$_{3}$)
and (\ref{10.2d}) are zero at $k_{x}=0$ and at $\omega ,~k_{x}=0$,
respectively. Based on the forms of $\mathbf{K}_{J}\left( \omega
^{2},k_{x}\right) $ as generated by the Magnus expansion, and evident from %
\eqref{9.2} for the leading order contributions, it may be demonstrated that %
\eqref{-1}$_{2}$ is consistent with imaginary $\mathbf{S}_{\mathbf{m}}$ that
is zero at $\omega =0$. 
It is noteworthy that keeping the density $\pmb{\rho }^{\left( \mathrm{eff%
}\right) }$ as $\left\langle \rho \right\rangle $ or as any other scalar is
generally not possible since this would contradict the pure dynamic term on
the right-hand side of (\ref{10.2d}). Finally, it is emphasized that, by
virtue of (\ref{9.1}), the stress-impulse tensor defined as a pure dynamic
quantity $S_{ijk}=S_{ijk}\left( \omega \right) $ vanishes in the case of a
unit cell with any symmetric heterogeneity profile $\mathbf{Q}\left( 
\widetilde{y}\right) $ (regarding the "inaccessible" part $\mathbf{S}_{%
\mathbf{m}}-\mathbf{S}_{\mathbf{m}}^{+},$ see \S \ref{dis1}).

Equations (\ref{10.2}) and \eqref{-1} with polynomials $\mathbf{K}_{J}\left(
\omega ^{2},k_{x}\right) $ given by the Magnus series (\ref{8}) imply that the elastic moduli $c_{ijkl}^{( \mathrm{eff} ) }$ are rational
functions of $\omega ^{2},k_{x}$ while the density and coupling terms 
$\pmb{\rho }^{( \mathrm{eff} )) }$, $\omega S_{ijk}$ are functions of $\omega ^{2}$,  defined by the series 
\begin{equation} \label{10.3}
\big\{ 
\mathbf{c}^{( \mathrm{eff} ) }(\omega ^{2},k_{x}),\, 
\pmb{\rho }^{( \mathrm{eff} ) }(\omega ^{2}),\, 
\mathbf{S}(\omega ) \big\} = 
\big\{ \mathbf{c}^{\left( 0\right) },\, \langle\rho \rangle \mathbf{I},\, 0
 \big\} +
\sum\limits_{m=1,2,...}
\big\{ 
\mathbf{c}^{\left( m\right) },\, 
\pmb{\rho }^{\left( 2m\right) },\, 
\mathbf{S}^{\left(2m-1\right) }
 \big\} ,
\end{equation}%
with real $\pmb{\rho }^{\left( m\right) }$ and imaginary $S_{ijk}^{\left(
m\right) }$ proportional to $\omega ^{m}$, $c_{ijkl}^{\left( m\right) }$
real or imaginary depending as $m$ is odd or even, respectively. These
series are similar to (\ref{8}) in that they are majorised by the power
series in long-wave parameter $\varepsilon .$ The Magnus series with $M$
terms enables finding $M$ terms of the series (\ref{10.3}). 
It is apparent from (\ref{10.2}$_{1}$), (\ref{-1}$_{1}$) and \eqref{1.1}
that $\mathbf{S}_{\mathbf{n}}$ and $\pmb{\rho }^{( \mathrm{eff} )) }$ depend only upon $\mathbf{N}_{2}$ and $\rho $, thus 
\begin{align}
\mathbf{S}_{\mathbf{n}}\left( \omega \right) 
& =-i\omega \left\langle \mathbf{N}_{2}\right\rangle ^{-1}\big\{\mathbf{a}%
_{2}^{\left( 1\right) }+\omega ^{2}\big(\mathbf{a}_{3}^{\left( 3\right) }-%
\mathbf{a}_{4}^{\left( 2\right) }\left\langle \mathbf{N}_{2}\right\rangle
^{-1}\mathbf{a}_{2}^{\left( 1\right) }\big)+\ldots \big\},  \label{13}
\\
\pmb{\rho }^{( \mathrm{eff} ) }(\omega )& =\left\langle \rho
\right\rangle \mathbf{I}-\omega ^{2}\big(\mathbf{a}_{7}^{\left( 2\right) }-%
\mathbf{a}_{2}^{\left( 1\right) }\left\langle \mathbf{N}_{2}\right\rangle
^{-1}\mathbf{a}_{2}^{\left( 1\right) }\big)+\ldots ,  \notag 
\end{align}%
with {\small 
\begin{align}
\mathbf{a}_{2}^{\left( 1\right) }& =\frac{T}{2}\int_{0}^{1}\int_{0}^{%
\widetilde{\varsigma }}\left( \rho \mathbf{N}_{2}-\mathbf{N}_{2}\rho \right)
\ ( =\mathbf{a}_{2}^{\left( 1\right) \mathrm{T}}) ,  \notag
\label{13.1} \\
\big\{ \mathbf{a}_{4}^{\left( 2\right) }
,\ \mathbf{a}_{7}^{\left( 2\right) } ( =\mathbf{a}_{7}^{\left( 2\right) \mathrm{T}})
\big\} 
& =\frac{T^{2}}{6}\int_{0}^{1}\int_{0}^{%
\widetilde{\varsigma }}\int_{0}^{\widetilde{\varsigma }_{1}}
\big\{ 
2\mathbf{N}_{2}\rho \mathbf{\mathbf{N}}_{2}-\mathbf{N}_{2}\mathbf{\mathbf{N}}_{2}\rho
-\rho \mathbf{N}_{2}\mathbf{\mathbf{N}}_{2} 
,  \ 
 \rho \rho \mathbf{N}_{2}+\mathbf{\mathbf{N}}_{2}\rho \rho 
-2\rho \mathbf{N}_{2}\rho 
\big\},
\\
\mathbf{a}_{3}^{\left( 3\right) }& =\frac{T^{3}}{6}\int_{0}^{1}\int_{0}^{%
\widetilde{\varsigma }}\int_{0}^{\widetilde{\varsigma }_{1}}\int_{0}^{%
\widetilde{\varsigma }_{2}}\left( 2\rho \mathbf{N}_{2}\rho \mathbf{N}_{2}-2%
\mathbf{N}_{2}\rho \mathbf{N}_{2}\rho +\mathbf{N}_{2}\mathbf{N}_{2}\rho \rho
-\rho \rho \mathbf{N}_{2}\mathbf{N}_{2}\right) ,\   \notag
\end{align}%
} in which $\mathrm{d}\widetilde{\varsigma },~\mathrm{d}\widetilde{\varsigma 
}_{1},...$ are suppressed (as kept tacit hereafter) and dependence of
co-factors on the successive integration variables $\widetilde{\varsigma },~%
\widetilde{\varsigma }_{1},...$ is understood. The remaining part of the
coupling tensor is only obtainable through the combination $\mathbf{S}_{%
\mathbf{m}}+\mathbf{S}_{\mathbf{m}}^{+}$, and it depends upon $\mathbf{N}%
_{1} $, $\mathbf{N}_{2}$ and $\rho $, with 
\begin{align}
\mathbf{S}_{\mathbf{m}}+\mathbf{S}_{\mathbf{m}}^{+}& =i\omega \big(%
\left\langle \mathbf{N}_{1}\right\rangle ^{\mathrm{T}}\left\langle \mathbf{N}%
_{2}\right\rangle ^{-1}\mathbf{a}_{2}^{\left( 1\right) }-\mathbf{a}%
_{2}^{\left( 1\right) }\left\langle \mathbf{N}_{2}\right\rangle
^{-1}\left\langle \mathbf{N}_{1}\right\rangle ^{\mathrm{T}}-\mathbf{a}%
_{5}^{\left( 1\right) }\big)+\text{O}(\omega ^{3}),\ \ 
\text{where} \\
\mathbf{a}_{5}^{\left( 1\right) }& =\frac{1}{2}T\int_{0}^{1}\int_{0}^{%
\widetilde{\varsigma }}\big((\mathbf{N}_{1}-\mathbf{N}_{1}^{\mathrm{T}})\rho
-\rho (\mathbf{N}_{1}-\mathbf{N}_{1}^{\mathrm{T}})\big)\  ( =-\mathbf{a}%
_{5}^{\left( 1\right) \mathrm{T}}) .  \notag
\end{align}

For example, the expansions of the inertial quantities for the transversely
isotropic layered medium discussed in \S \ref{expan} are $\pmb{\rho }^{(%
\mathrm{eff})}=\left\langle \rho \right\rangle \mathbf{I}+\pmb{\rho }%
^{(2)}+\ldots $ {\small 
\begin{align}
\rho _{22}^{(2)}& =(\omega T)^{2}\bigg\{\frac{1}{6}\int_{0}^{1}\int_{0}^{%
\widetilde{\varsigma }}\int_{0}^{\widetilde{\varsigma }_{1}}\big(\rho \rho
c_{22}^{-1}+c_{22}^{-1}\rho \rho -2\rho c_{22}^{-1}\rho \big)-\langle
c_{22}^{-1}\rangle ^{-1}\big(\frac{1}{2}\int_{0}^{1}\int_{0}^{\widetilde{%
\varsigma }}(c_{22}^{-1}\rho -\rho c_{22}^{-1})\big)^{2}\bigg\},  \notag
\label{-8} \\
S_{222}^{(1)}& =\frac{i}{2}\omega T\langle c_{22}^{-1}\rangle
^{-1}\int_{0}^{1}\int_{0}^{\widetilde{\varsigma }}(c_{22}^{-1}\rho -\rho
c_{22}^{-1}), \\
S_{112}^{(1)}& =\frac{i}{2}\omega T\bigg\{\langle \frac{c_{12}}{c_{22}}%
\rangle ^{-1}\langle c_{22}^{-1}\rangle ^{-1}\int_{0}^{1}\int_{0}^{%
\widetilde{\varsigma }}(c_{22}^{-1}\rho -\rho
c_{22}^{-1})+\int_{0}^{1}\int_{0}^{\widetilde{\varsigma }}\bigg(\big(1-\frac{%
c_{12}}{c_{22}}\big)\rho -\rho \big(1-\frac{c_{12}}{c_{22}}\big)\bigg)\bigg\}%
,  \notag
\end{align}%
} and $\rho _{11}^{(2)}$, $S_{211}^{(1)}=S_{233}^{(1)}$ have respectively
the same form as $\rho _{22}^{(2)}$, $S_{222}^{(1)}$ with $c_{22}$ replaced
by $c_{66}$ in \eqref{-8}$_{1,2}$.

\subsubsection{Effective medium defined from the Floquet dispersion}

\label{secfloq}

Modelling a dispersive effective medium may be based on a more relaxed
approach that abandons fitting the matrix $i\mathbf{K}$ to the coefficients
of sextic system of wave equations and deals instead with the asymptotic
secular equation for the eigenvalues $iK_{\alpha }$ or $e^{iK_{\alpha }T}$
of $i\mathbf{K}$ or $\mathbf{M}\left( T,0\right) $, which is a dispersion
equation for the onset of fundamental Floquet branches $K_{\alpha }\left(
\omega ,k_{x}\right) $ or $\omega _{\alpha }\left( k_{x},K\right) $ analyzed
in \cite{N,N1,N2}.  This gives the same secular equation as that for the $i\mathbf{K}$ matrices, and hence  preserves the long-wave Floquet dispersion but not the 
  displacement-traction   vector $\mathbf{w}_{\alpha }$  at the period edges (see (\ref{4.1}%
)).  By not fitting all of the physical properties, this type of approach to  homogenization modelling  introduces extra degrees of  freedom.  In particular, a "modified" effective medium may be defined that is 
asymptotically similar to  $i\mathbf{K}$  but the matrix $\widetilde{\mathbf{Q}}_{%
\mathrm{eff}}$ has no pure dynamic terms in the diagonal blocks,  and
hence matches the Stroh-like form (\ref{1.1}), (\ref{1.2}) (though now with (%
\ref{12.1})$_{1}$\textbf{),} i.e. satisfies the standard form of the
governing equations (\ref{0}) with dispersive effective coefficients.

For instance, in the 1D case $k_{x}=0,$ the matrix 
\begin{equation}
\mathbf{Q}_{\mathrm{eff}}=\left\langle \mathbf{Q}\right\rangle +i\mathbf{K}%
^{\left( 1\right) }+i\mathbf{K}^{\left( 2\right) }=i%
\begin{pmatrix}
i\omega ^{2}\mathbf{a}_{2}^{\left( 1\right) } & \left\langle \mathbf{N}%
_{2}\right\rangle +\omega ^{2}\mathbf{a}_{4}^{\left( 2\right) } \\ 
-\left\langle \rho \right\rangle \omega ^{2}\mathbf{I}+\omega ^{4}\mathbf{a}%
_{7}^{\left( 2\right) } & -i\omega ^{2}\mathbf{a}_{2}^{\left( 1\right) }%
\end{pmatrix}
\label{14}
\end{equation}%
has asymptotically (to the order of this matrix itself) the same secular
equation as the matrix 
\begin{equation}
\widetilde{\mathbf{Q}}_{\mathrm{eff}}=i%
\begin{pmatrix}
\mathbf{0} & \left\langle \mathbf{N}_{2}\right\rangle +\omega ^{2}\mathbf{a}%
_{4}^{\left( 2\right) } \\ 
-\left\langle \rho \right\rangle \omega ^{2}\mathbf{I}+\omega ^{4}(\mathbf{a}%
_{7}^{\left( 2\right) }-\mathbf{a}_{2}^{(1)2}\left\langle \mathbf{N}%
_{2}\right\rangle ^{-1}) & \mathbf{0}%
\end{pmatrix}%
.  \label{15}
\end{equation}%
The latter "skips" (by construction) the Willis coupling tensor and leads to
the same definition of the matrix of second-order elastic coefficients $%
\left( nn\right) ^{\left( 2\right) }$ as in (\ref{10.2}), while the
second-order density matrix $\widetilde{\pmb{\rho }}^{\left( 2\right)
}=\omega ^{2}(\mathbf{a}_{2}^{(1)2}\left\langle \mathbf{N}_{2}\right\rangle
^{-1}-\mathbf{a}_{7}^{\left( 2\right) })$ following from (\ref{15}) is
generally different from $\pmb{\rho }^{\left( 2\right) }$ in (\ref{13})
due to non-commutativity of $\left\langle \mathbf{N}_{2}\right\rangle $ and $%
\mathbf{a}_{2}^{(1)}.$ See also the SH example in Appendix B.

\section{Effective medium coefficients for SH waves}

\label{SHsec}

\subsection{The wave number matrix}

Consider SH waves in an isotropic medium with periodic density $\rho \left(
y\right) $ and shear modulus $\mu \left( y\right) .$ The SH state vector $%
\mathbf{\eta }\left( y\right) =\left( A,iF\right) ^{\mathrm{T}},$ where $A$
and $F$ are the amplitudes of $u=u_{3}$ and $\sigma _{23}$ (the indices
correspond to $\mathbf{u}\parallel X_{3},\ \mathbf{n}\parallel X_{2},\ 
\mathbf{m}\parallel X_{1}$), satisfies Eq. (\ref{1}) with the system matrix%
\begin{equation}
\mathbf{Q}\left( y\right) =i%
\begin{pmatrix}
0 & -\mu ^{-1} \\ 
\mu k_{x}^{2}-\rho \omega ^{2} & 0%
\end{pmatrix}%
.  \label{SH1}
\end{equation}%
The 2$\times $2 case leads to some simplifications not available for higher
algebraic dimensions. In particular, the two eigenvalues of the monodromy
matrix $\mathbf{M}(T,0),$ which are the inverse of one other (since $\det 
\mathbf{M}=1$ due to the isotropy), are defined by the single quantity $%
\mathrm{tr\,}\mathbf{M}(T,0)$. The implications are explored in \cite{SKN}
and only the necessary equations are cited here - the reader is referred to 
\cite{SKN} for details. The main result is that the wave number matrix, and
hence the effective system matrix 
$\mathbf{Q}_{\mathrm{eff}}\left( \omega \right) =i\mathbf{K}$ has semi-explicit form, 
\begin{equation}
\mathbf{Q}_{\mathrm{eff}}  =i 
\begin{pmatrix}
K_{1} & K_{2} \\ 
K_{3} & -K_{1}%
\end{pmatrix}%
=\frac{K}{\sin KT}\left[ \mathbf{M}(T,0)-
\mathbf{I}\, \cos KT\right] ,  
\ \ KT=\cos^{-1} \big(\frac{1}{2}\mathrm{tr\,}\mathbf{M}(T,0)\big),
\label{N4.0}
\end{equation}
where Re$\,\cos^{-1} \in \lbrack 0,\pi ]$, Im$\,\cos^{-1} \geq 0,$ and $\pm K$
(no subscript) are the eigenvalues of $\mathbf{K.}$ 

\subsection{Willis equations and effective coefficients}

Following the general formalism of \S \ref{sec3} the effective material is
assumed to have constitutive equations described by the Willis model, which
in this case has only a single momentum component $p_{3}$ and the usual
stress components for SH waves in elasticity.  Noting that $S_{53}= 0$, on account of the  transversely isotropic axis $\mathbf{n}$, we have 
\begin{equation}  
\begin{pmatrix}
\sigma _{13} \\ 
\sigma _{23} \\ 
p_{3}%
\end{pmatrix}%
=%
\begin{pmatrix}
c_{55}^{(\mathrm{eff})} & c_{54}^{(\mathrm{eff})} & 0 \\ 
c_{45}^{(\mathrm{eff})} & c_{44}^{(\mathrm{eff})} & S_{43} \\ 
0 & S_{43} & \rho ^{(\mathrm{eff})}%
\end{pmatrix}%
\begin{pmatrix}
u_{,1} \\ 
u_{,2} \\ 
\dot{u}%
\end{pmatrix}%
.  \label{331}
\end{equation}

\noindent These constitutive relations imply, using (\ref{11}$_{1}$), that
the governing equation for the SH displacement is of the form%
\begin{equation}
c_{44}^{( \mathrm{eff} ) }u^{\prime \prime }+\big(\omega ^{2}\rho
^{( \mathrm{eff} ) }-k_{x}^{2}c_{55}^{( \mathrm{eff} ) }%
\big)u=0,  \label{SH6}
\end{equation}%
where $^{\prime }$ means \textrm{d}$/$\textrm{d}$y$. The coupling term $%
S_{43}$ is absent from the equation of motion, as expected
from the Willis equations \eqref{11} for a scalar problem. At the same time,
(\ref{331}) leads to the state-vector system matrix in the form%
\begin{equation}
\mathbf{Q}_{\mathrm{eff}}=i%
\begin{pmatrix}
-{c_{44}^{(\mathrm{eff})}}^{-1}(k_{x}c_{45}^{(\mathrm{eff})}-\omega S_{43})
& -{c_{44}^{(\mathrm{eff})}}^{-1} \\ 
&  \\ 
k_{x}^{2}c_{55}^{(\mathrm{eff})}-\omega ^{2}\rho ^{(\mathrm{eff})}+{c_{44}^{(%
\mathrm{eff})}}^{-1}(k_{x}c_{45}^{(\mathrm{eff})}-\omega S_{43})^{2} & ~{%
c_{44}^{(\mathrm{eff})}}^{-1}(k_{x}c_{45}^{(\mathrm{eff})}-\omega S_{43})%
\end{pmatrix}%
,  \label{--=}
\end{equation}%
where $c_{45}^{(\mathrm{eff})+}=c_{54}^{(\mathrm{eff})}=-c_{45}^{(\mathrm{eff%
})}$ has been used. 


Setting $\mathbf{Q}_{\mathrm{eff}}$ of the Willis model equal to that of %
\eqref{N4.0} gives the material parameters 
\begin{align}
c_{44}^{(\mathrm{eff})}& =-K_{2}^{-1}, 
\quad
 \rho ^{(\mathrm{eff})}=-\omega ^{-2}\big(K_{3}(0)+K_{2}^{-1}(0)K_{1}^{2}(0)\big),  
\quad
S_{43}=-\omega
^{-1}K_{2}^{-1}(0)K_{1}(0),  \label{SH}  
\notag \\
c_{55}^{(\mathrm{eff})}& =k_{x}^{-2}\big(K_{3}+K_{2}^{-1}K_{1}^{2}+\omega
^{2}\rho ^{(\mathrm{eff})}\big), 
\quad  
c_{45}^{(\mathrm{eff})} =k_{x}^{-1}\big(%
K_{2}^{-1}K_{1}-K_{2}^{-1}(0)K_{1}(0)\big),    
\end{align}%
with $K_{J}(0)=K_{J}(\omega ,0)$. These may be expressed directly in terms
of the elements of the monodromy matrix, using the form \eqref{N4.0} along
with $\det \mathbf{K}=-K^{2}$, 
\begin{align} 
c_{44}^{(\mathrm{eff})} &=\frac{\sin KT}{iKM_{2}}, 
\quad 
\rho ^{(\mathrm{eff})} =\frac{K^{2}(0)}{\omega ^{2}}c_{44}^{(\mathrm{eff})}(0),  
\quad
S_{43} =\frac{M_{4}(0)-M_{1}(0)}{2\omega M_{2}(0)}, 
\notag  \label{=0}
\\
c_{55}^{(\mathrm{eff})}& =k_{x}^{-2}\big(\omega ^{2}\rho ^{(\mathrm{eff})}-K^{2}c_{44}^{(\mathrm{eff})}\big), 
\quad
c_{45}^{(\mathrm{eff})} =k_{x}^{-1}\big(\omega S_{43}+\frac{M_{1}-M_{4}}{%
2M_{2}}\big),  
\end{align}
where $M_{J}=M_{J}(T,0)$ are functions of $\omega $ and $k_{x}$, and $(0)$
means evaluated at $k_{x}=0$. Note that the expressions for $\rho ^{(\mathrm{%
eff})}$ and $c_{55}^{(\mathrm{eff})}$ also follow from the equation of
motion \eqref{SH6} and its solution $u(y)=u(0)e^{iKy}$, using $k_{x}=0$ for $%
\rho ^{(\mathrm{eff})}(\omega )$.

Explicit expressions for the low-frequency long-wave expansion of the
material parameters may be found in the same manner as in \S %
\ref{expan} for the general case. The starting point is the Magnus expansion 
$\mathbf{Q}_{\mathrm{eff}}=\left\langle \mathbf{Q}\right\rangle +i\mathbf{K}%
^{\left( 1\right) }+i\mathbf{K}^{\left( 2\right) }$ for the SH wave number
matrix. Details of the analysis and a summary of the results are presented
in Appendix B.

\subsection{Examples and discussion}

\subsubsection{A bilayered unit cell}

The general formulation is illustrated by the case of a two-component
piecewise constant unit cell. Specifically, consider a periodic structure of
homogeneous isotropic layers $j=1,2$, each with constant density $\rho _{j},$
shear modulus $\mu _{j}$ and thickness $d_{j}.$ The monodromy matrix $%
\mathbf{M}\left( T,0\right) =\mathrm{e}^{\mathbf{Q}_{2}d_{2}}\mathrm{e}^{%
\mathbf{Q}_{1}d_{1}}\equiv \mathbf{M}\left( \omega ,k_{x}\right) $ has the
well-known form 
\begin{equation}
\mathbf{M}\left( \omega ,k_{x}\right) =%
\begin{pmatrix}
\cos \psi _{2}\cos \psi _{1}-\frac{\gamma _{1}}{\gamma _{2}}\sin \psi
_{2}\sin \psi _{1} & -\frac{i}{\gamma _{1}}\cos \psi _{2}\sin \psi _{1}-%
\frac{i}{\gamma _{2}}\sin \psi _{2}\cos \psi _{1} \\ 
-i\gamma _{1}\cos \psi _{2}\sin \psi _{1}-i\gamma _{2}\sin \psi _{2}\cos
\psi _{1} & \cos \psi _{2}\cos \psi _{1}-\frac{\gamma _{2}}{\gamma _{1}}\sin
\psi _{2}\sin \psi _{1}%
\end{pmatrix}%
,  \label{152}
\end{equation}%
where $\psi _{j}=d_{j}\sqrt{\mu _{j}^{-1}\rho _{j}\omega ^{2}-k_{x}^{2}}$ is
the phase shift over a layer 
and $\gamma _{j}=\mu _{j}\psi _{j}/d_{j}$, see \cite{SKN}. 
\begin{figure}[th]
\begin{center}
\includegraphics[width=5in,height=3in]{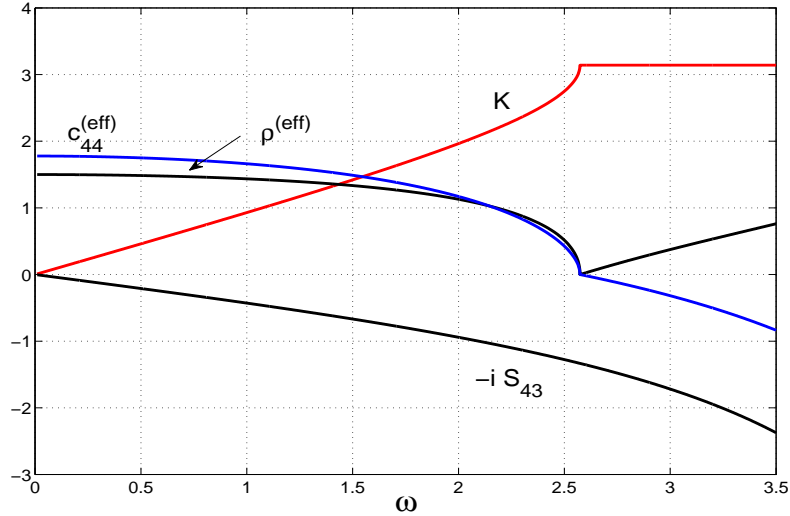}
\end{center}
\caption{The effective material properties of the bilayered SH case for $%
k_{x}=0$: elastic moduli, inertial parameters and the effective wave number
are plotted in blue, black and red, respectively. The frequency range
includes the first band edge which is at the frequency where Re$\,K=\protect%
\pi $ first occurs. Only the real parts of the quantities indicated are
plotted. For frequencies in the stop band the imaginary parts of $c_{44}^{(%
\mathrm{eff})}$, $\protect\rho ^{(\mathrm{eff})}$ and $K$ are non-zero but
not shown. }
\label{fig1}
\end{figure}
Figures \ref{fig1} and \ref{fig2} show the computed parameters for the case
of layers of equal thickness, $d_{1}=d_{2}=1/2$, with $\rho _{1}=1$, $%
c_{1}=1 $; $\rho _{2}=2$, $c_{2}=2$, where $c_{j}$ is the shear wave speed $%
(c^{2}=\mu /\rho )$.  
Figure \ref{fig1}  shows the effective parameters for propagation normal to
the layers ($k_{x}=0$). The vanishing of both $c_{44}^{(\mathrm{eff})}$ and $\rho ^{(%
\mathrm{eff})}$ at the band edge at $\omega =\omega _{1}\approx 2.6$ is
expected on the basis of the fact that $\mathbf{Q}_{\mathrm{eff}}$ is
singular at the band edge and scales as $(\omega -\omega _{1})^{-1/2}$ near
it \cite{SKN}. Referring to the 12-element in Eq. \eqref{--=}, this implies
first that $c_{44}^{(\mathrm{eff})}\propto (\omega -\omega _{1})^{1/2}$ and
then, from the 21-element and the finite value of $\det \mathbf{Q}_{\mathrm{%
eff}}$, that $\rho ^{(\mathrm{eff})}\propto (\omega -\omega _{1})^{1/2}$.
The square root decay of both $c_{44}^{(\mathrm{eff})}$ and $\rho ^{(\mathrm{%
eff})}$ is apparent in Figure \ref{fig1}.


\begin{figure}[th]
\begin{center}
\includegraphics[width=6in]{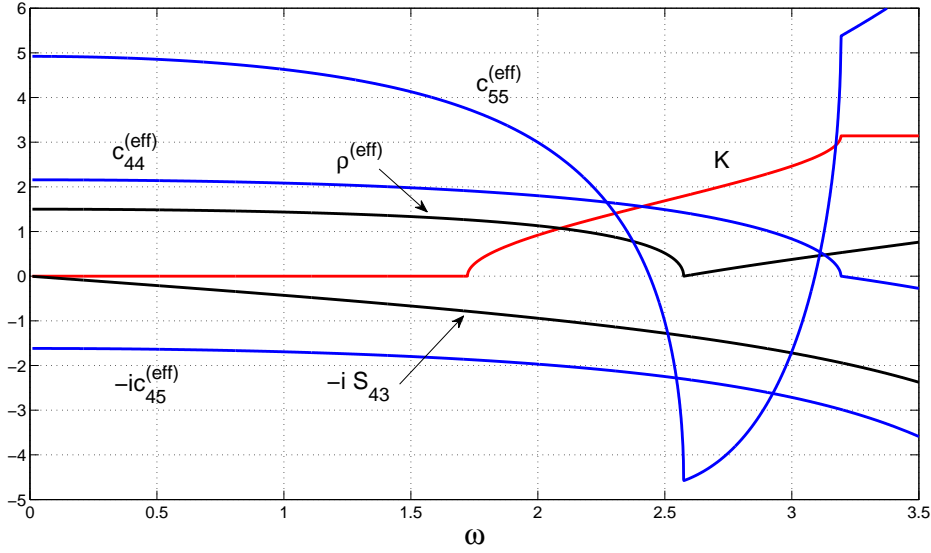}
\end{center}
\caption{The same as in Figure \protect\ref{fig1} but for $k_{x}=1.0$. The
additional parameters $c_{55}^{(\mathrm{eff})}$ and $c_{45}^{(\mathrm{eff})}$
are relevant to $k_{x}\neq 0$. Only the real parts of the quantities are
plotted.}
\label{fig2}
\end{figure}

The wavenumber is finite, $k_{x}=1$, in Figure \ref{fig2}. This has the
effect of increasing the frequency of the band edge, and introducing a range
of frequency from $\omega =0$ up to the cut-on at $\omega \approx 1.7$ in
which the effective wave is non-propagating. Note that $\rho ^{(\mathrm{eff}%
)}$ and $S_{43}$ are unchanged from Figure \ref{fig1} while the elastic
modulus $c_{44}^{(\mathrm{eff})}$ is different, and tends to zero at the new
band edge as expected. The non-zero $k_{x}$ leads to non-zero $c_{45}^{(%
\mathrm{eff})}$, and the parameter $c_{55}^{(\mathrm{eff})}$ becomes
complex-valued at the $k_{x}=0$ band edge. Only the real parts of the
quantities are shown in both figures. No attempt is made here to discuss
their imaginary components, which requires careful analysis of the branch
cuts and is a topic for separate study.

\subsubsection{Reflection and transmission of a half-space of effective
material}

\label{45-}

\begin{figure}[th]
\begin{center}
\includegraphics[width=5in,height=3.in]{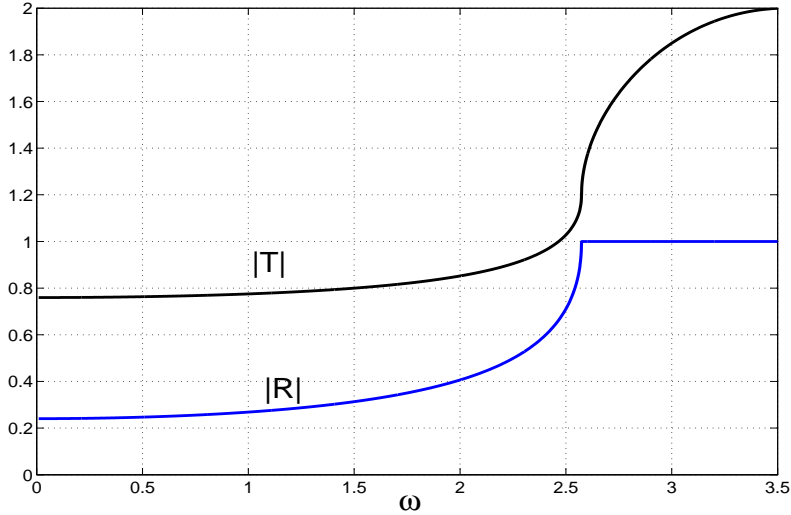}
\end{center}
\caption{ The magnitude of the reflection and transmission coefficients of 
\eqref{RT} for normal incidence $(\protect\theta =0).$}
\label{fig3}
\end{figure}

As an example of the type of boundary problem that can be solved using the
effective medium equations, consider reflection-transmission of SH waves at a
bonded interface $y=0$ between the half-space of the periodically stratified
medium ($y>0$) and a uniform half-space ($y<0$) of isotropic material with $%
\rho _{0}$, $\mu _{0}$ and $c_{0}=\sqrt{\mu _{0}/\rho _{0}}$. A SH plane
wave is incident from the uniform half-space with propagation direction at
angle $\theta $ from the interface normal. The total solution is taken as 
\begin{equation}
u(x,y)=e^{ik_{x}x}\times 
\begin{cases}
\big[e^{ik_{y}y}+Re^{-ik_{y}y}\big], & y\leq 0, \\ 
Te^{iK(\omega ,k_{x})y}, & y>0,%
\end{cases}%
\ \ \mathrm{with}\ (k_{x},\,k_{y})=\frac{\omega }{c_{0}}(\sin \theta ,\,\cos
\theta ).
\end{equation}%
The reflection and transmission coefficients $R$ and $T$ follow from the
continuity conditions for particle velocity and traction at the interface.
They may be expressed in standard form using  SH impedances defined
as $Z_{\pm }=-\sigma _{23}/\dot{u}|_{y=0_{\pm }}$. The impedance in the
uniform half-space is  $Z_{-}=\rho _{0}c_{0}\cos \theta .$
The impedance $Z_{+}$ for the effective medium follows from \eqref{331} as%
\begin{equation}
Z_{+}=\omega ^{-1}\big(Kc_{44}^{(\mathrm{eff})}+k_{x}c_{45}^{(\mathrm{eff}%
)}-\omega S_{43}\big). 
\end{equation}%
This is identical to  the impedance of the periodically stratified half-space because they both imply a ratio of components of the outgoing eigenvector $\mathbf{w}$ which is common to  $\mathbf{M}(T,0)$ and $\mathbf{K}$. 
In these terms, the continuity conditions for displacement and traction
yield the exact result%
\begin{equation}
1+R =T, \ \ 
Z_{-}(1-R) =Z_{+}T,
\qquad \Rightarrow \qquad R=\frac{Z_{-}-Z_{+}}{Z_{-}+Z_{+}},\ \ T=\frac{%
2Z_{-}}{Z_{-}+Z_{+}}.  \label{RT}
\end{equation}%
Figure \ref{fig3} shows $\vert R\left( \omega \right) \vert $ and 
$\vert T\left( \omega \right) \vert $ calculated for normal
incidence from a uniform half-space with $\rho _{1}=1$, $c_{1}=1$ on a
periodic structure of two layers with $\rho _{1}=1$, $c_{1}=1$, $\rho _{2}=2$%
, $c_{2}=2,$ which was used in Figs. 1 and 2. As expected, $|R|\leq 1$ with
total reflection in the stopband.

The explicit dependence of the reflection coefficient on the effective
medium parameters $c_{44}^{(\mathrm{eff})} $, $c_{45}^{(\mathrm{eff})} $ and 
$S_{43}$ means that, in principle, measurement of $R$ via experiment can
provide useful knowledge for their determination.

\subsubsection{Uniform normal impedance}

It is instructive to consider  the particular case of $k_{x}=0$ with 
$z=\sqrt{\rho \left( y\right) \mu \left( y\right) }=\rho \left( y\right)
c\left( y\right) $ independent of $y$, i.e. $z\equiv z_{0}$ . The 2$\times $2 
matrix $\mathbf{Q}\left( y\right) $   is then a scalar
multiple of a constant matrix, and  
$M_{1}(0)=M_{4}(0)$,  $M_{2}(0)=(i\omega z_{0})^{-1}\sin KT$ (where $\left( 0\right) $ stands
for $k_{x}=0$). As a result, by (\ref{SH}), the effective parameters at any $%
\omega $ retain their values obtained from static averaging: $S_{43}=0$, $%
c_{44}^{(\mathrm{eff})}(\omega ,0)=\langle \mu ^{-1}\rangle ^{-1}$ and $\rho
^{(\mathrm{eff})}(\omega )=\langle \rho \rangle $. This simplification is a
consequence of the fact that  constant  $z$ implies constant
eigenvectors of the SH matricant $\mathbf{M}\left( y,0\right) $ and hence no
reflection of SH waves normally propagating through a periodic structure,
which is in accordance with the physical meaning of the impedance  $z$.
Consistency is also observed in that the effective impedance $z^{(\mathrm{eff%
})}$ defined through the above effective parameters is equal to $z_{0},$ 
\begin{equation}
z^{(\mathrm{eff})}=\sqrt{\rho ^{(\mathrm{eff})}c_{44}^{(\mathrm{eff})}}=\rho
^{(\mathrm{eff})}c^{(\mathrm{eff})}=\sqrt{\langle \rho \rangle \langle \mu
^{-1}\rangle ^{-1}}=z_{0}.  \label{z}
\end{equation}%
The only effect of the inhomogeneity is to speed up or retard advancing
waves according to the effective speed $c^{(\mathrm{eff})}=z^{(\mathrm{eff}%
)}/\rho ^{(\mathrm{eff})}$, which in the present case follows from (\ref{z})
as $c^{(\mathrm{eff})}=z_{0}/\rho ^{(\mathrm{eff})}=\langle c^{-1}\rangle
^{-1}$.

 %

\subsubsection{Discussion}
In the case of purely uni-dimensional motion, $k_x=0$, the system \eqref{331} involves only the first three parameters of Eq. \eqref{=0}: 
$c_{44}^{(\mathrm{eff})}$, $\rho ^{(\mathrm{eff})}$, and $S_{43}$.  Willis \cite{Will}
derived expressions for the same  quantities for a laminated medium. His   coefficients  
\cite[Eq. (3.30)]{Will} relate   weighted means of strain and velocity
$(\langle we\rangle$, $\langle w\dot u \rangle)$ to ensemble means of   stress  and  momentum density $(\langle\sigma\rangle$, $\langle p\rangle)$, where $w$ is a general weighting function first introduced in  \cite{MW}, such that ensemble means correspond to $w=1$.   The Willis parameters derived here, e.g.  \eqref{331},  concern strain and velocity at the single point $y=0$ in the unit period, and therefore correspond to the specific weight function 
$w(x) = 2L \delta (x)$ in the notation of \cite{Will}.  It is important to note, however, that  the stress and momentum density used here are not ensemble averages but 
are quantities associated with the same point in the unit period.  This identification, for instance, means that the solution of  the reflection-transmission problem  of \S 
\ref{45-} is in fact the deterministic solution.    In summary, while the governing equations are the same in both cases, the Willis parameters developed here do not bear  a one-to-one correspondence with those in \cite{Will}.

\section{Conclusion}

\label{conc}

A fully dynamic homogenization scheme has been developed for periodically
layered anisotropic elastic solids. In the process,  the dispersive and nonlocal Willis model has been shown to provide an optimal constitutive setting for the effective medium.  The crucial point of
 the present method is the insistence that the matrix of
coefficients $\mathbf{Q}_{\mathrm{eff}}$ of the sextic system of
elastodynamics equations, for whatever homogeneous effective medium is
considered, must exactly match the Floquet wave number matrix $\mathbf{K}$
of the periodic system. This  is not a low-frequency long-wave
approach, so long as $\mathbf{K}$ is defined at the given frequency $\omega $ and
horizontal wave number $k_{x}$.  The wave number matrix 
$\mathbf{K=K}\left( \omega ,k_{x}\right) $ is an  analytic function of $\omega $ and $k_{x}$
that may be explicitly defined via the Magnus series expansion, which is guaranteed to converge
below the first Floquet stopband at the edge of the Brillouin zone. The
choice of constitutive model for the effective medium is critical.  We have
demonstrated that the standard anisotropic elasticity theory does not
suffice as it cannot provide a $\mathbf{Q}_{\mathrm{eff}}$ to properly account  
for dynamic terms appearing in the wave number matrix $\mathbf{K}\left(
\omega ,k_{x}\right) $.  On the other hand, the Willis model for the  effective
medium, which includes coupling effects, can allow us to associate elements
of the effective system matrix $\mathbf{Q}_{\mathrm{eff}}$ with elements in $%
\mathbf{K}$. The main results are contained in Eqs. \eqref{10.2} and %
\eqref{-1} which infer the material parameters of the effective Willis
medium from $\mathbf{K.}$ Invoking the Magnus series, explicit expressions
for the low-frequency long-wave expansion of these effective Willis
parameters have been found and the accuracy for their truncated asymptotics
has been estimated.

The example of SH plane wave reflection and transmission considered in \S %
\ref{45-} indicates the type of application possible using the Willis
effective medium. The point is not so much to provide new solutions for
layered media, although it is simpler to formulate and solve such problems
using equations for a homogeneous model. The potential power of the dynamic
effective medium model is that it is possible to relate the effective
properties of the Willis material to measurable dynamic quantities. Thus,
the reflection and transmission problem illustrates how the reflection
coefficient $R$ depends on a certain combination of the Willis parameters.
Measurements of $R=R(\omega ,k_{x})$ provide a means to characterize
periodic layered systems as equivalent homogeneous but dispersive materials.
Other problems that may be considered are, for instance, surface wave
propagation in a periodically layered half space, waveguides comprised of
periodic layers, and point forces.

\subsubsection*{Acknowledgements}

This work has been supported by the grant ANR-08-BLAN-0101-01 from the ANR
(Agence Nationale de la Recherche) and by the project SAMM (Self-Assembled
MetaMaterials) from the cluster AMA (Advanced Materials in Aquitaine).
A.N.N. is grateful to the Laboratoire de M\'{e}canique Physique (LMP) of the
Universit\'{e} Bordeaux 1 for the hospitality. 


\section*{Appendix}

\subsection*{A. Estimates for the long-wavelength expansion of $\mathbf{K}$}

\subsubsection*{A.1 Auxiliary notations}

For any $2d\times 2d$ matrix $\mathbf{A}$ consisting of $d\times d$ blocks $%
\mathbf{A}_{J}$ ($J=1,...,4$), denote a 2$\times $2 matrix of matrix norms $%
\Vert \mathbf{A}_{J}\Vert $ by $\Vert \mathbf{A}\Vert_{2\times 2}:$ 
\begin{equation}
\big\Vert \mathbf{A}\big\Vert _{2\times 2}=%
\begin{pmatrix}
\Vert \mathbf{A}_{1}\Vert & \Vert \mathbf{A}_{2}\Vert
\\ 
\Vert \mathbf{A}_{3}\Vert & \Vert \mathbf{A}_{4}\Vert%
\end{pmatrix}%
.  \label{M0}
\end{equation}%
Note that $\Vert \mathbf{AB}\Vert _{2\times 2}\leq \Vert 
\mathbf{A}\Vert _{2\times 2}\Vert \mathbf{B}\Vert _{2\times
2},$ where and hereafter a matrix inequality is understood as that between
the corresponding matrix elements. Let%
\begin{equation}
\mathbb{\nu }^{2}=\left\langle N_{1}\right\rangle ^{2}+\left\langle
N_{2}\right\rangle \left\langle N_{3}\right\rangle ,\ 
q\left( y\right)
=\max_{\ }\left( {\Vert \mathbf{N}_{J}\left( y\right) \Vert 
}/{\langle N_{J}\rangle },\ {\rho \left( y\right) }/{%
\left\langle \rho \right\rangle }\right) ,  \label{M1}
\end{equation}%
where $\left\langle \cdot \right\rangle $ is the averaging symbol 
and $\left\langle N_{J}\right\rangle $ is an average of a norm $%
N_{J}\left( y\right) =\Vert \mathbf{N}_{J}\left( y\right) \Vert $
of the Stroh-matrix block ($J=1,2,3$)$.$ ($\Vert \mathbf{N}%
_{1}\Vert =\Vert \mathbf{N}_{4}\Vert $)$.$ Both $\mathbb{%
\nu }$ and $q\left( y\right) $ are physically dimensionless and strictly
positive. The magnitude of $\mathbb{\nu }$ is usually of the order of 1
unless a high-contrast case affecting the averaged profile; in turn, the
averaged value $\left\langle q\right\rangle ,$ strictly speaking, satisfies $%
1\leq \left\langle q\right\rangle \leq 4$ but is typically close to 1 as
well. Let us define the long-wave small parameter $\varepsilon $ as 
\begin{equation}
\varepsilon \equiv kT\ \mathrm{with}\ k=\sqrt{k_{x}^{2}+
 {\omega ^{2}}/{V^{2}}},\ \ V^{2}=
\big(\left\langle N_{1}\right\rangle ^{2}+\left\langle
N_{3}\right\rangle \left\langle N_{2}\right\rangle \big)\big( \left\langle \rho
\right\rangle \left\langle N_{2}\right\rangle \big)^{-1}.  \label{7}
\end{equation}%
It is noted that $\epsilon =\mathbb{\nu }\varepsilon $ may equally be taken
as such a parameter and that it can actually be replaced everywhere below by
a smaller value 
\begin{equation}
\widetilde{\epsilon }=T\max \left( k_{x}\left\langle N_{1}\right\rangle
,k_{x}\sqrt{\left\langle N_{2}\right\rangle \left\langle N_{3}\right\rangle }%
,\omega \sqrt{\left\langle \rho \right\rangle \left\langle
N_{2}\right\rangle }\right) <\mathbb{\nu }\varepsilon .  \label{7.1}
\end{equation}%
Next, define%
\begin{equation}
\mathbf{C}= 
\begin{pmatrix}
\mathbf{I} & \mathbf{0} \\ 
\mathbf{0} & \frac{T\left\langle N_{2}\right\rangle }{\mathbb{\nu }%
\varepsilon }\mathbf{I}%
\end{pmatrix}%
  ,\ \widehat{\mathbf{Q}}\left( y\right) =\mathbf{CQ}\left( y\right) 
\mathbf{C}^{-1}=i 
\begin{pmatrix}
k_{x}\mathbf{N}_{1} & \frac{\mathbb{\nu }\varepsilon }{T\left\langle
N_{2}\right\rangle }\mathbf{N}_{2} \\ 
\frac{T\left\langle N_{2}\right\rangle }{\mathbb{\nu }\varepsilon }\left(
k_{x}^{2}\mathbf{N}_{3}-\rho \omega ^{2}\right) & k_{x}\mathbf{N}_{1}^{%
\mathrm{T}}%
\end{pmatrix}%
 ,  \label{M2}
\end{equation}%
where $\mathbf{Q}\left( y\right) $ is given in (\ref{1.1}), and the
normalization of $\widehat{\mathbf{Q}}\left( y\right) $ 
provides a common estimate $\big\Vert \widehat{\mathbf{Q}}_{J}\left(
y\right) \big\Vert T\leq \mathbb{\nu }\varepsilon q\left( y\right) $ for
the blocks of $\widehat{\mathbf{Q}},$ from which follow the estimates for $%
\big\Vert \mathbf{Q}_{J}\left( y\right) \big\Vert .$ Denoting the
(constant) 2$\times $2 matrices 
\begin{equation}
\mathbf{H}= 
\begin{pmatrix}
1 & 1 \\ 
1 & 1%
\end{pmatrix}%
  ,\ \big\Vert \mathbf{C}\big\Vert _{2\times 2}\equiv \widetilde{%
\mathbf{C}}= 
\begin{pmatrix}
1 & 0 \\ 
0 & \frac 1{\mathbb{\nu }\varepsilon }{T\left\langle N_{2}\right\rangle }%
\end{pmatrix}%
  ,\ \mathbf{\Omega }=\frac{1}{T}\mathbb{\nu }\varepsilon \left\langle
q\right\rangle \widetilde{\mathbf{C}}^{-1}\mathbf{H}\widetilde{\mathbf{C}}%
,  \label{M3}
\end{equation}%
and using the notation (\ref{M0}) enables us to write the blockwise
estimates for $\widehat{\mathbf{Q}}\left( y\right) $ and for $\mathbf{Q}%
\left( y\right) $ in the form%
\begin{equation}
\mathbf{\ }%
\begin{array}{c}
\big\Vert \widehat{\mathbf{Q}}\left( y\right) \big\Vert _{2\times 2}T\leq 
\mathbb{\nu }\varepsilon q\left( y\right) \mathbf{H\ }\Rightarrow \
\big\Vert \mathbf{Q}\left( y\right) \big\Vert _{2\times 2}=\widetilde{%
\mathbf{C}}^{-1}\big\Vert \widehat{\mathbf{Q}}\left( y\right) \big\Vert
_{2\times 2}\widetilde{\mathbf{C}}\leq \frac{q\left( y\right) }{\left\langle
q\right\rangle }\mathbf{\Omega } \\ 
\mathbf{\ }\Rightarrow \big\Vert \left\langle \mathbf{Q}\right\rangle
\big\Vert _{2\times 2}\leq \left\langle \big\Vert \mathbf{Q}\big\Vert
_{2\times 2}\right\rangle \leq \mathbf{\Omega .}%
\end{array}
\label{M4}
\end{equation}%
Note the meaning of $\mathbf{\Omega }$ as a matrix of upper bounds of blocks
of the statically averaged system matrix $\left\langle \mathbf{Q}%
\right\rangle $.

\subsubsection*{A.2 Estimates for the Magnus series $i\mathbf{K}%
=\sum_{m=0}^{\infty }i\mathbf{K}^{\left( m\right) }$}

An elegant proof that the Magnus series (\ref{8}) converges for $%
\left\langle \big\Vert \mathbf{Q}\big\Vert _{2}\right\rangle T<\pi $ \cite%
{MN} is somewhat implicit in that it does not provide fully explicit
estimates of the series terms and remainder. These may be obtained for a
narrower range by adapting the derivation detailed and referenced in \cite%
{BCOR}. It proceeds from the estimate%
\begin{equation}
\Vert \mathbf{K}^{\left( m\right) }\Vert T  \leq \pi \left( \xi T 
\left\langle \Vert \mathbf{Q}\Vert \right\rangle \right)
^{m+1},\  \ 
\xi =2\big( \int_{0}^{\pi } [2+x( 1-\cot x ) ]^{-1} \mathrm{d}x
\big)^{-1}
=1.8400...  \label{M5}
\end{equation}%
with specifically the matrix norm $\Vert \cdot \Vert _{2}$ as
kept tacit below. (Note aside that the physical dimension of $\left( \mathbf{%
Q}T\right) ^{n}$ is the same for any $n$ as that of $\mathbf{Q}T$ and $%
\mathbf{K}^{\left( m\right) }T$.) Let $i\widehat{\mathbf{K}}%
=\sum_{m=0}^{\infty }i\widehat{\mathbf{K}}^{\left( m\right) }$ be the same
series (\ref{8}) but related to the matrix $\widehat{\mathbf{Q}}$ in place
of $\mathbf{Q}$. Applying (\ref{M5}) yields the blockwise estimates for
series terms $\widehat{\mathbf{K}}^{\left( m\right) }$ and, hence, $\mathbf{K%
}^{\left( m\right) }$ in the form 
\begin{equation}
\Vert \widehat{\mathbf{K}}^{\left( m\right) }\Vert _{2\times
2}T\leq \pi X^{m+1}\mathbf{H\ }\Rightarrow \ \Vert \mathbf{K}^{\left(
m\right) }\Vert _{2\times 2}\leq \pi \xi X^{m}\mathbf{\Omega },
\ \ \text{where } \  X=\xi \mathbb{\nu }\varepsilon \left\langle q\right\rangle .
\label{M6}
\end{equation}%
Assume hereafter that $X<1$ (which is within the convergence radius $X<\xi \frac{\pi}{2} $ that follows from the result $\left\langle \big\Vert \mathbf{Q}%
\big\Vert \right\rangle T<\pi $ of \cite{MN}). By (\ref{M6}), the residual
series $\widehat{\mathbf{R}}^{\left( M\right) }=\sum_{m=M}^{\infty }i%
\widehat{\mathbf{K}}^{\left( m\right) }$ and $\mathbf{R}^{\left( M\right)
}=\sum_{m=M}^{\infty }i\mathbf{K}^{\left( m\right) }$ satisfy%
\begin{equation}
\Vert \widehat{\mathbf{R}}^{\left( M\right) }\Vert _{2\times
2}T\leq \pi (1-X )^{-1}  X^{M+1} \mathbf{H\ }\Rightarrow \ \Vert \mathbf{R%
}^{\left( M\right) }\Vert _{2\times 2}\leq \pi \xi (1-X )^{-1} X^{M} %
\mathbf{\Omega ,}  \label{M7}
\end{equation}%
so that they decrease as $M$ grows and tend to zero as $M\rightarrow \infty
. $

Knowing the upper bound of $ \Vert \mathbf{R}^{\left( M\right)
} \Vert _{2\times 2}$ evaluates the sufficient number of terms to be
kept in the Magnus series to ensure a desired accuracy of truncation for a
fixed long-wave parameter $\varepsilon ,$ or else provides the value of $%
\varepsilon $ that ensures this accuracy for a given truncation step. The
accuracy is gauged by the matrix $\mathbf{\Omega }$ of blockwise bounds of $%
\left\langle \mathbf{Q}\right\rangle ,$ see (\ref{M4}). For example, let the
Magnus series (\ref{8}) be truncated as $i\mathbf{K}=\left\langle \mathbf{Q}%
\right\rangle +i\mathbf{K}^{\left( 1\right) }+i\mathbf{K}^{\left( 2\right) }$
and the remainder $\mathbf{R}^{\left( 3\right) }$ discarded. According to (%
\ref{M7})$_{2},\  \Vert \mathbf{R}^{\left( 3\right) } \Vert
_{2\times 2}\leq 5.8X^{3}/\left( 1-X\right) .$ Thus taking the spectral
range as $\mathbb{\nu }\varepsilon \left\langle q\right\rangle <0.24$ or $%
<0.128$ ensures $ \Vert \mathbf{R}^{\left( 3\right) } \Vert
_{2\times 2}<\mathbf{\Omega }$ or $<0.1\mathbf{\Omega },$ respectively (note
that truncating $i\mathbf{K}$ by $\left\langle \mathbf{Q}\right\rangle $ at $%
\mathbb{\nu }\varepsilon \left\langle q\right\rangle <0.128$ discards $%
 \Vert \mathbf{R}^{\left( 1\right) } \Vert _{2\times 2}<1.79\mathbf{%
\Omega }$).

{
It is noted that the above mentioned sufficient criterion for the
Magnus series convergence and the bounds for its terms restrict $\omega $
and $k_{x}$ by  imposing conditions on the norm 
$\left\Vert \mathbf{Q}\right\Vert $ of 
$\mathbf{Q=Q}\left( y;\omega ,k_{x}\right)$.  At the same
time, if $\mathbf{Q}$ is independent of $y$ then the Magnus series certainly
converges 
at any $\omega $ and $k_{x}.$  Hence if the
inhomogeneity is relatively weak so that $\left\Vert \mathbf{Q-}\left\langle 
\mathbf{Q}\right\rangle \right\Vert $ is markedly smaller than $\left\Vert 
\mathbf{Q}\right\Vert ,$ then the $\omega $ and $k_{x}$ bounds on the Magnus
series can be eased by using a different 
approach that is based, instead of (\ref{M5}), on the estimate 
$\left\Vert \mathbf{K}^{\left( m\right) }\right\Vert \leq Cm\left\langle
\left\Vert \mathbf{Q-}\left\langle \mathbf{Q}\right\rangle \right\Vert
\right\rangle ^{m}$ ($C>0$ is some constant).   The latter acquires a growing factor $m$ but is hinged explicitly on  $\left\Vert \mathbf{Q-}\left\langle \mathbf{Q}\right\rangle \right\Vert $, which is why this approach can be 
 much more advantageous for small 
$\left\Vert \mathbf{Q-}\left\langle \mathbf{Q}\right\rangle \right\Vert $.
}

\subsubsection*{A.3 Invertibility of $\mathbf{K}_{2}$}

Derivation of the effective material constants requires inverting the block $%
\mathbf{K}_{2}$ (see \S 3.2). Evidently $i\mathbf{K}_{2}=i\left\langle 
\mathbf{N}_{2}\right\rangle +\mathbf{R}_{2}^{\left( 1\right) }$ is assuredly
negative definite (like $\left\langle \mathbf{N}_{2}\right\rangle $) if 
$
 \Vert \left\langle \mathbf{N}_{2}\right\rangle ^{-1}\mathbf{R}%
_{2}^{\left( 1\right) } \Vert <1$.  
Inserting $ \Vert \mathbf{R}_{2}^{\left( 1\right) } \Vert \leq \pi
\xi \left\langle q\right\rangle \left\langle N_{2}\right\rangle X/\left(
1-X\right) $ from (\ref{M7})$_{2}$ and resolving the resulting inequality
with respect to $\mathbb{\nu }\varepsilon \left\langle q\right\rangle $
yields%
\begin{equation}
\mathbb{\nu }\varepsilon \left\langle q\right\rangle <
\big( 
2+11\left\langle q\right\rangle \left\langle N_{2}\right\rangle  \Vert
\left\langle \mathbf{N}_{2}\right\rangle ^{-1} \Vert \big)^{-1}.  
\label{M9}
\end{equation}%
This condition can be further improved by using more precise estimates for
the low-order terms of the Magnus series. For this purpose, it is suitable
to proceed from $\widehat{\mathbf{K}}^{\left( m\right) }$ defined by (\ref{8}%
) with $\widehat{\mathbf{Q}}.$ Using $\int_{0}^{1}...\int_{0}^{\widetilde{%
\varsigma }_{m-1}}q\left( \widetilde{\varsigma }\right) ...q\left( 
\widetilde{\varsigma }_{m}\right) \mathbf{H}^{m}=\frac{2^{m-1}\left\langle
q\right\rangle ^{m}}{m!}\mathbf{H}$ and $ \Vert \mathbf{K}_{2}^{\left(
m\right) } \Vert =\frac{T\left\langle N_{2}\right\rangle }{\mathbb{\nu }%
\varepsilon } \Vert \widehat{\mathbf{K}}_{2}^{\left( m\right)
} \Vert $ gives for $m=1,2:$ 
\begin{equation}
\begin{array}{c}
 \Vert \widehat{\mathbf{K}}^{\left( 1\right) } \Vert _{2\times
2}T\leq \frac{\left( 2\mathbb{\nu }\varepsilon \left\langle q\right\rangle
\right) ^{2}}{4}\mathbf{H,\ } \Vert \widehat{\mathbf{K}}^{\left(
2\right) } \Vert _{2\times 2}T\leq \frac{\left( 2\mathbb{\nu }%
\varepsilon \left\langle q\right\rangle \right) ^{3}}{9}\mathbf{H\Rightarrow 
} \\ 
 \Vert \mathbf{K}_{2}^{\left( 1\right) } \Vert \leq \mathbb{\nu }%
\varepsilon \left\langle q\right\rangle \left\langle N_{2}\right\rangle ,\
 \Vert \mathbf{K}_{2}^{\left( 2\right) } \Vert \leq \frac{4}{9}%
\mathbb{\nu }\varepsilon \left\langle q\right\rangle ^{2}\left\langle
N_{2}\right\rangle ,%
\end{array}
\label{M10}
\end{equation}%
where the right-hand sides are smaller than in (\ref{M6}) with $m=1,2.$
Exploiting (\ref{M10}) leads to 
\begin{equation}
\begin{array}{c}
 \Vert \left\langle \mathbf{N}_{2}\right\rangle ^{-1}\mathbf{R}%
_{2}^{\left( 1\right) } \Vert = \Vert \left\langle \mathbf{N}%
_{2}\right\rangle ^{-1}\left( \sum_{m=1}^{2}i\mathbf{K}_{2}^{\left( m\right)
}+\mathbf{R}^{\left( 3\right) }\right)  \Vert \\ 
\leq 2\left\langle q\right\rangle \left\langle N_{2}\right\rangle  \Vert
\left\langle \mathbf{N}_{2}\right\rangle ^{-1} \Vert \left(
0.272X+0.132X^{2}+{2.93X^{3}}/(1-X)\right) \\ 
\Rightarrow  \Vert \left\langle \mathbf{N}_{2}\right\rangle ^{-1}\mathbf{%
R}_{2}^{\left( 1\right) } \Vert <1\ \mathrm{if\ }X<\min \left( 0.41,%
\big( {1+1.4\left\langle q\right\rangle \left\langle N_{2}\right\rangle
 \Vert \left\langle \mathbf{N}_{2}\right\rangle ^{-1} \Vert }\big)^{-1}%
\right) .%
\end{array}
\label{M11}
\end{equation}%
Hence, by (\ref{M11}), taking a typical value $\left\langle
q\right\rangle \left\langle N_{2}\right\rangle  \Vert \left\langle 
\mathbf{N}_{2}\right\rangle ^{-1} \Vert \approx 1$ guarantees that $%
\mathbf{K}_{2}$ is invertible in the spectral range $X<0.41$, i.e. $\mathbb{%
\nu }\varepsilon \left\langle q\right\rangle <0.22.$

\subsubsection*{A.4 Estimate for the eigenvalues of $\mathbf{K}$}

The upper bound for eigenvalues $K_{\alpha }$ of $\mathbf{K}$ 
may be evaluated via $ \Vert K_{\alpha } \Vert \leq  \Vert 
\mathbf{K} \Vert $; however, taking note that $K_{\alpha }$ are also
the eigenvalues of $\widehat{\mathbf{K}}$ yields a better estimate with
regard for (\ref{M7})$_{1}$ as follows%
\begin{equation}
 \Vert K_{\alpha } \Vert \leq  \Vert \widehat{\mathbf{K}}%
 \Vert = \Vert \widehat{\mathbf{R}}^{\left( 0\right) } \Vert
\leq 2 \Vert \widehat{\mathbf{R}}^{\left( 0\right) } \Vert_{2\times 2} \leq 2\pi T^{-1}(1-X)^{-1}X 
\label{M12}
\end{equation}%
(note that $K_{\alpha }$ here may certainly take either real or complex
values). For instance, $ \Vert K_{\alpha } \Vert T<0.62$ in the
long-wave range $\mathbb{\nu }\varepsilon \left\langle q\right\rangle <0.128$
($X<0.236$) that was shown in \S A.2 to ensure accuracy of truncation of $%
\mathbf{K}$ after $\mathbf{K}^{\left( 2\right) }.$ Note that vanishing of
the right-hand side of (\ref{M12}) at $X=0,$ i.e. at $\varepsilon =T\sqrt{%
k_{x}^{2}+\omega ^{2}/V^{2}}=0$ (see (\ref{7})) is in agreement with 
$\mathbf{K}= 
\begin{pmatrix}
\mathbf{0} & \langle \mathbf{N}_{2}\rangle \\ 
\mathbf{0} & \mathbf{0}%
\end{pmatrix}$ 
 at $ \omega =0,\ k_{x}=0$.  

\subsubsection*{A.5 Convergence of $\partial \mathbf{K}/\partial k_{x}$}

Consider $\left( \partial \mathbf{K}/\partial k_{x}\right) _{k_{x}=0}\equiv 
\mathbf{K}^{\prime }\left( 0\right) $ as defined by Eq. (\ref{++2}). The
series (\ref{++1}$_{1}$) and (\ref{++2}$_{1}$) assuredly converge if so does
the same series for $\ln \mathbf{CMC}^{-1}=\mathbf{C}\left( \ln \mathbf{M}%
\right) \mathbf{C}^{-1},$ where $\mathbf{C}$ is introduced in (\ref{M2}$_{1}$%
). In turn the series for $\ln \mathbf{CMC}^{-1}$ assuredly converges if 
\begin{equation}
\big\Vert \mathbf{CMC}^{-1}\mathbf{-I}\big\Vert \leq \big\Vert \exp
\langle \Vert \widehat{\mathbf{Q}}\Vert \rangle 
\mathbf{-I}\big\Vert \leq \exp \left( 2\mathbb{\nu }\varepsilon
\left\langle q\right\rangle \right) -1<1\ \Rightarrow \ 2\mathbb{\nu }%
\varepsilon \left\langle q\right\rangle <\ln 2,  \label{M14}
\end{equation}%
where it was used that $\big\Vert \widehat{\mathbf{Q}}\big\Vert =2\mathbb{%
\nu }\varepsilon \left\langle q\right\rangle $ due to (\ref{M4}$_{1}$) and
 $\Vert \mathbf{H}\Vert =2.$ For $\mathbf{M}\equiv \mathbf{M}%
\left( 0\right) $ with $k_{x}=0,$ which is the case in hand, the condition (%
\ref{M14}) reduces, using (\ref{7.1}), to 
\begin{equation}
2\widetilde{\epsilon }\left( 0\right) \left\langle q\left( 0\right)
\right\rangle <\ln 2,  \label{M15}
\end{equation}%
where$\ \widetilde{\epsilon }\left( 0\right) =T\max ( \omega \sqrt{%
\left\langle \rho \right\rangle \left\langle N_{2}\right\rangle }) ,\
q\left( 0\right) =\max_{\ }\big( {\Vert \mathbf{N}_{2}\left(
y\right) \Vert }/{\left\langle N_{2}\right\rangle },{\rho \left(
y\right) }/{\left\langle \rho \right\rangle }\big) \ \left( 1\leq
\left\langle q\left( 0\right) \right\rangle \leq 2\right) .$

The integrals in (\ref{++1}$_{2}$) and (\ref{++2}$_{2}$) exist provided the
matrix $\mathbf{I}+ x (\mathbf{M}(0)-\mathbf{I})$ is invertible for any $x\in (
0,1 )$. According to \cite{MN}, this is guaranteed if 
\begin{equation}
2\widetilde{\epsilon }\left( 0\right) \left\langle q\left( 0\right)
\right\rangle <\pi ,  \label{M16}
\end{equation}%
which implies that the eigenvalues $e^{iK_{\alpha }\left( 0\right) T}$ of $%
\mathbf{M}\left( 0\right) $ do not attain real negative values. This
condition coincides with the sufficient condition for the convergence of
Magnus series at $k_{x}=0$.

\subsection*{B. Effective medium coefficients for SH waves at long wavelength%
}

According to (\ref{8}) truncated by $\mathbf{K}^{\left( 2\right) }$, 
\begin{equation}
\mathbf{Q}_{\mathrm{eff}} =i 
\begin{pmatrix}
K_{1}^{\left( 1\right) } & -\left\langle \mu ^{-1}\right\rangle
+K_{2}^{\left( 2\right) } \\ 
\left\langle \mu \right\rangle k_{x}^{2}-\left\langle \rho \right\rangle
\omega ^{2}+K_{3}^{\left( 2\right) } & -K_{1}^{\left( 1\right) }%
\end{pmatrix}
,  \label{SH2}
\end{equation}%
where $K_{2,3}^{\left( 1\right) }=0,\ K_{1}^{\left( 2\right) }=0$ and
(omitting the integration variables){\small 
 \begin{align} 
K_{1}^{\left( 1\right) }&=\frac{iT}{2}\int_{0}^{1}\int_{0}^{\widetilde{%
\varsigma }}\left[ \mu ^{-1}, \rho \omega ^{2}-\mu k_{x}^{2} \right] , 
\notag \\
\big\{ 
K_{2}^{( 2) } 
,\ 
K_{3}^{( 2) } 
\big\} 
&=\frac{T^{2}}{3}\int_{0}^{1}\int_{0}^{\widetilde{%
\varsigma }}\int_{0}^{\widetilde{\varsigma }_{1}} 
\big\{ 
\left[ \mu ^{-1}, \left[
\rho \omega ^{2}-\mu k_{x}^{2} ,\mu ^{-1}\right] \right] 
, \ 
\left[ \rho \omega
^{2}-\mu k_{x}^{2} ,\left[ \mu ^{-1}, \rho \omega ^{2}-\mu k_{x}^{2} \right] %
\right] 
\big\} 
.    \label{SH3}
\end{align}
}On the other hand, Eq. (\ref{--=}) taken to the same order as (\ref{SH2}%
) and written with Voigt index notation, yields 
\begin{equation}
\mathbf{Q}_{\mathrm{eff}}=i 
\begin{pmatrix}
- \left\langle \mu ^{-1}\right\rangle \left( k_{x}c_{45}^{\left( 1\right)
}-\omega S_{43}^{\left( 1\right) }\right) & -\left\langle \mu
^{-1}\right\rangle +c_{44}^{\left( 2\right) }\left\langle \mu
^{-1}\right\rangle ^{2} \\ 
\begin{array}{c}
k_{x}^{2}\left( \left\langle \mu \right\rangle +c_{55}^{\left( 2\right)
}\right) -\omega ^{2}\left( \left\langle \rho \right\rangle +\rho ^{\left(
2\right) }\right) \\ 
+\left\langle \mu ^{-1}\right\rangle \left( k_{x}c_{45}^{\left( 1\right)
}-\omega S_{43}^{\left( 1\right) }\right) ^{2}%
\end{array}
& \left\langle \mu ^{-1}\right\rangle \left( k_{x}c_{45}^{\left( 1\right)
}-\omega S_{43}^{\left( 1\right) }\right)%
\end{pmatrix}
,  \label{SH4}
\end{equation}%
where $c_{45}^{( 1 )+} = c_{54}^{( 1 )} = - c_{45}^{( 1 )}$ has been used,
and in addition $c_{44}^{\left( 0\right) }=\langle \mu
^{-1}\rangle ^{-1},\ c_{55}^{\left( 0\right) }=\left\langle \mu
\right\rangle $, $c_{44}^{\left( 1\right) }=c_{55}^{\left( 1\right) }=0,\
\rho ^{\left( 1\right) }=0$ with, as noted 
earlier, $S_{53}^{\left( 1\right) } = 0$. From (\ref{SH2})-(\ref{SH4}), 
\begin{equation}
\begin{array}{c}
c_{45}^{\left( 1\right) }=\frac{i}{2}k_{x}T \left\langle
\mu^{-1}\right\rangle^{-1} \int_{0}^{1}\int_{0}^{\widetilde{\varsigma }}%
\left[ \mu ^{-1},\mu \right] ,\ S_{43}^{\left( 1\right) }= \frac{i}{2}\omega
T \left\langle \mu^{-1}\right\rangle^{-1} \int_{0}^{1}\int_{0}^{\widetilde{%
\varsigma }}\left[ \mu ^{-1},\rho \right] ,\  \\ 
\ c_{44}^{\left( 2\right) }=\left\langle \mu ^{-1}\right\rangle
^{-2}K_{2}^{\left( 2\right) },\ k_{x}^{2}c_{55}^{\left( 2\right) }-\rho
^{\left( 2\right) }\omega ^{2}=K_{3}^{\left( 2\right) }- \left\langle \mu
^{-1}\right\rangle^{-1 } K_{1}^{\left( 1\right)2} ,%
\end{array}
\label{SH5}
\end{equation}%
which identifies $c_{45}^{\left( 1\right) }\left( k_{x}\right) =-
c_{54}^{\left( 1\right) }\left( k_{x}\right) $,\ $c_{44}^{\left( 2\right)
}\left( \omega ^{2},k_{x}^{2}\right)$ and $S_{43}^{\left( 1\right) }(\omega)$%
. The 
density correction $\rho ^{\left( 2\right) }(\omega)$ follows from the final
identity as equal to the expression \eqref{-8}$_1$ with $c_{33}\rightarrow
\mu$, 
and $c_{55}^{\left( 2\right) }\left(\omega^2, k_{x}^{2}\right) $ is then
uniquely defined. Similar results follow for the SH waves in a monoclinic
periodic medium.

Note that the secular equation of the SH-wave matrix $\mathbf{Q}_{\mathrm{eff%
}}$ given by (\ref{SH2}) can equally be associated, up to the same order as $%
\mathbf{Q}_{\mathrm{eff}}$ itself, with other matrices such as, e.g., the
matrix 
\begin{equation}
\widetilde{\mathbf{Q}}_{\mathrm{eff}}=i 
\begin{pmatrix}
0 & -\left\langle \mu ^{-1}\right\rangle +K_{2}^{\left( 2\right) } \\ 
\left\langle \mu \right\rangle k_{x}^{2}-\left\langle \rho \right\rangle
\omega ^{2}+K_{3}^{\left( 2\right) }- \left\langle \mu
^{-1}\right\rangle^{-1 } {K_{1}^{\left( 1\right) 2}} & 0%
\end{pmatrix}
,  \label{SH7}
\end{equation}%
which has zero diagonal components like in (\ref{SH1}) and thus leads to
Eqs. (\ref{331}), (\ref{SH5}) but with zero $c_{45}^{\left( 1\right) }$ and $%
S_{43}^{\left( 1\right) }.$ However, as remarked in \S \ref{secfloq}, the
effective polarization of the SH displacement-traction eigenmodes is
asymptotically defined by the eigenvectors $\mathbf{w}_{1,2}$ of (truncated) 
$\mathbf{Q}_{\mathrm{eff}}=i\mathbf{K,}$ while its definition from the
eigenvectors $\widetilde{\mathbf{w}}_{1,2}$ of $\widetilde{\mathbf{Q}}_{%
\mathrm{eff}}$\textbf{\ }is different since $\widetilde{\mathbf{w}}%
_{1,2}\neq \mathbf{w}_{1,2}$ to the first order in\textbf{\ }$\varepsilon $. 
\textbf{\medskip }

\pagebreak


\end{document}